\documentclass[10pt]{elsarticle0}
\usepackage{epsfig}
\usepackage{slashed}
\usepackage{amsmath}
\usepackage{amssymb}
\newcommand{\beqn}{\begin{eqnarray}}
\newcommand{\eeqn}{\end{eqnarray}}

\usepackage{verbatim} % adds environment for commenting out blocks of text & for better verbatim
\usepackage{amsmath}
\usepackage{amssymb}
\usepackage{braket}
\usepackage{color}

\newcommand{\be}{\begin{eqnarray}}
\newcommand{\ee}{\end{eqnarray}}

\newcommand{\rev}[1]{{#1}}
\newcommand{\rv}[1]{{#1}}
\newcommand{\rvv}[1]{{#1}}

\newcommand{\zz}[1]{{#1}}
\newcommand{\cor}[1]{{#1}}

\begin{document}

\title{Topological invariant in terms of the Green functions for the Quantum Hall Effect in the presence of varying magnetic field}

\author[ARIEL]{M.A.~Zubkov
\footnote{
Corresponding author,
e-mail: zubkov@itep.ru, On leave of absence from NRC "Kurchatov Institute" - ITEP, B. Cheremushkinskaya 25, Moscow, 117259, Russia}
 }
\author[ARIEL]{Xi Wu}

\address[ARIEL]{ Physics Department, Ariel University, Ariel 40700, Israel}

%\address[ICM]{Instituto de Ciencia de Materiales de Madrid,
%CSIC, Cantoblanco; 28049 Madrid, Spain.}

%\address[MIPT]{Moscow Institute of Physics and Technology, 9, Institutskii per., Dolgoprudny, Moscow Region, 141700, Russia}

%\address[DFU]{Far Eastern Federal University,  School of Biomedicine, 690950 Vladivostok, Russia}

\begin{abstract}
Recently the Wigner - Weyl formalism has been applied to the lattice models of solid state physics and to the lattice regularized quantum field theory. This allows to demonstrate that the electric current of intrinsic Anomalous Quantum Hall effect is expressed through the momentum space topological invariant composed of the Green functions both for the two - and the three - dimensional systems. Here we extend this consideration to the  case of the Quantum Hall Effect existing in the presence of arbitrarily varying external magnetic field. The corresponding electric current appears to be proportional to the topological invariant in phase space composed of the Wigner transformed Green function that depends both on coordinates and momenta.
\end{abstract}

%%%%%%%%%%%%%%%%%%%%%%%%%%%%%%%%%%%%%%%%%%%%%%%%%%%%%%%%%%%%%%%%%%%%%%%%

%%%%%%%%%%%%%%%%%%%%%%%%%%%%%%%%%%%%%%%%%%%%%%%%%%%%%%%%%%%%%%%%%%%%%%%

\maketitle

\section{Introduction}

Relation between topology and the Quantum Hall effect including the Anomalous Quantum Hall Effect (AQHE) follows the discovery of the so - called TKNN invariant  \cite{TTKN,TKNN0,Haldane}: the Hall conductivity appears to be proportional to the integral of Berry curvature over the occupied branches of spectrum. The corresponding three - dimensional constructions were discussed, for example, in \cite{3ddiophantine,Hall3DTI}.  This formalism allows to calculate the Quantum Hall Effect (QHE) of electrons in the two - dimensional periodic potentials (or in the 2D lattice models) in the presence of constant magnetic field \cite{Fradkin}. However, it does not give the transparent calculation method for the QHE of Bloch electrons in the presence of varying external magnetic field. Besides, although by construction the TKNN invariant is defined for the noninteracting system, it is well - known that the total AQHE conductivity is robust to the introduction of disorder and weak interactions. The representation of this conductivity as the topological invariant composed of the Green functions was proposed for the two dimensional system in \cite{Matsuyama:1986us,Volovik0} (for more details see Chapter 21.2.1 in \cite{Volovik2003}, see also \cite{So,Golterman1993,IshikawaMatsuyama1986}). Using this approach the interacting system may be considered: the full two - point Green function of the interacting system should be substituted to the corresponding expression. It is usually believed, that the higher order full Green functions do not give a contribution to the QHE although there is still no direct proof of this statement. Such topological invariants were further discussed in the series of papers (see, for example, \cite{Gurarie2011,EssinGurarie2011} and references therein). In \cite{Z2016_1,Z2016_2} the similar relation between Hall conductivity and the topological invariants composed of the Green functions was deduced for the three - dimensional systems with the AQHE (in particular, for the discovered recently Weyl semimetals \cite{semimetal_effects10,semimetal_effects11,semimetal_effects12,semimetal_effects13,Zyuzin:2012tv,tewary}).
The role of disorder in the QHE has been discussed within several different approaches (see \cite{Fradkin,TKNN2,QHEr,Tong:2016kpv} and references therein). However, the introduction of disorder within the formalism of \cite{Matsuyama:1986us,Volovik0,Volovik2003,Z2016_1,Z2016_2} remains problematic: the corresponding topological invariants formally describe the idealized systems without impurities, where the Hall current remains in the bulk. Disorder pushes the Hall current to the boundary. But its total amount integrated over all space remains given by the topological invariants discussed in \cite{Matsuyama:1986us,Volovik0,Volovik2003,Z2016_1,Z2016_2}. In the present paper we will present the alternative proof of this conjecture for the case of the sufficiently weak disorder. Moreover, we extend the consideration of \cite{Z2016_1} from the AQHE to the case of the Quantum Hall Effect in the presence of arbitrarily varying external magnetic field. Technically its consideration is more complicated than that of the AQHE in periodic systems without external magnetic field because the two - point Green function ${\cal G}(p_1,p_2)$ now depends on two momenta $p_1,p_2$ instead of one unlike the case of the AQHE.
We hope, that the topological invariants proposed in the present paper may become useful for the consideration of the models with essential inhomogeneity since they are composed of the Green functions and inherit the beautiful mathematical structure of the topological invariants of \cite{Matsuyama:1986us,Volovik0,Volovik2003,Gurarie2011,EssinGurarie2011,Z2016_1}.
It is worth mentioning that the Hall conductance is believed to be robust to the introduction of weak interactions while strong interactions may bring the system to the phase with the fractional QHE (see, for example, \cite{Tong:2016kpv}). Its consideration, however, remains out of the scope of the present paper. For the general review of the topological aspects of the QHE we refer  to  \cite{Fradkin,Tong:2016kpv,Hatsugai} and references therein.

We explore the Wigner - Weyl formalism \cite{1,2,berezin,6} developed in \cite{Z2016_1,Z2016_2} for the lattice models of condensed matter physics (as well as for the lattice regularized quantum field theory). For the ordered review of this formalism see \cite{SZ2018}.
We demonstrate that the Hall conductance integrated over the whole space is given by the topological invariant in phase space (that consists of both coordinates and momenta). This topological invariant is expressed through the Wigner transformation of the two - point Green functions. Its structure repeats that of the topological invariants discussed in \cite{Matsuyama:1986us,Volovik0,Volovik2003,Gurarie2011,EssinGurarie2011,Z2016_1,Z2016_2} with the ordinary product substituted by the star product and the extra integration over the whole space added \footnote{The topological invariants of \cite{Matsuyama:1986us,Volovik0,Volovik2003,Gurarie2011,EssinGurarie2011,Z2016_1,Z2016_2} have the same structure as the degree of mapping of the three - dimensional manyfold to a group of matrices.}.  The value of our topological invariant responsible for the Hall conductance is not affected by the sufficiently weak disorder. We also find indications that it remains robust to the Electromagnetic interactions with sufficiently small effective coupling constant.

The paper is organized as follows. In Sect. \ref{Sect2} we introduce the notations that are useful for the investigation of various lattice models in momentum space. In Sect. \ref{Sect3} we recall the application of the Wigner - Weyl formalism to the lattice models proposed in \cite{Z2016_1,Z2016_2}. This formalism results, in particular, in the expression for the current of the AQHE through the topological invariants in momentum space. In Sect. \ref{Sect4} we present the extension of this result to the case of the lattice model in the presence of varying external magnetic field. Our main result is the expression for the QHE current through the topological invariant in phase space given in Eqs. (\ref{calM0}), (\ref{JA2D0}), (\ref{calM2d2}). In Sect. \ref{Sect5} we check the obtained results applying our expressions to the consideration of the problem with known solution - the QHE in the noninteracting system in the presence of constant external magnetic field. In Sections \ref{Sect2} -- \ref{Sect5} we consider the idealized systems without disorder and with the interactions neglected.  In Sect. \ref{SectConcl} we end with the conclusions, and discuss briefly the role of disorder. For the discussion of interactions we refer to \cite{ZZ2019_2}.

In Appendixes we accumulate various results used in the main text of the paper. The majority of those results are not original and were proved in the other publications. We feel this important to collect the derivation of those results in the Appendixes to our present paper for completeness.

\section{The description of the tight - binding models in momentum space}

\label{Sect2}

Following \cite{Z2016_1,Z2016_2,SZ2018} we start from the lattice tight - binding fermionic model with the partition function of the following form
\rev{\begin{equation}
Z=\int D\bar{\Psi}D\Psi  {\rm exp}\left(\sum_{x,y}\bar\Psi^T(x)\left(-i\mathcal{D}_{x,y}\right)\Psi(y)\right)\,.
\label{Z00}
 \end{equation}}
Here $\mathcal{D}_{x,y}$ is a matrix that depends on the discrete lattice coordinates $x$, $y$. $\Psi,\bar{\Psi}$ are the multi - component Grassmann - valued fields defined on the lattice sites. The corresponding indices are omitted here and below for brevity. \rvv{This partition function describes the {\it noninteracting} fermionic quasiparticles. Interactions result in the terms in exponent that are proportional to the higher powers of $\bar{\Psi}$ and $\Psi$. Here we do not discuss them and bound ourselves by the noninteracting model. The role of Coulomb interactions between electrons is considered in details in \cite{ZZ2019_2}, where it has been shown that the expression for the Hall conductivity proposed in the present paper does not receive corrections from interactions to any order in the effective fine structure constant. The noninteracting partition function} may be rewritten in momentum space as follows
\rev{\begin{eqnarray}
Z &=& \int D\bar{\psi}D\psi \, {\rm exp}\Big(  \int_{\cal M} \frac{d^D {p}}{|{\cal M}|}\nonumber\\&&\bar{\psi}^T({p}){Q}(p)\psi({p}) \Big)\,,\label{Z01}
\end{eqnarray}}
where integration is over the fields defined in momentum space $\cal M$. $|{\cal M}|$  is its volume, $D$ is the dimensionality of space - time, $\bar{\psi}$ and $\psi$ are the anticommuting multi - component Grassmann variables defined in momentum space. Without loss of generality we assume that time is discretized, so that momentum space is compact, and its volume is finite. In condensed matter physics time typically is not discretized,  the corresponding partition function may be obtained easily as the limit of Eq. (\ref{Z00}) when the time spacing tends to zero. The partition function of Eq. (\ref{Z01}) allows to describe the non - interacting fermion systems corresponding to matrix  $Q(p)$ (that is the Fourier transform of the lattice tight - binding matrix $D_{x y}$). The meaning of $Q(p)$ for the lattice models of electrons in crystals is the inverse propagator of Bloch electron.  \rvv{As it was mentioned above, } the interactions are not taken into account at this stage. Their effect as well as the effect of disorder will be discussed later. \rvv{In the present paper we concentrate on the case of vanishing temperature.}

Introduction of the external gauge field $A(x)$ defined as a function of coordinates effectively leads to the Peierls substitution (see, for example, \cite{Z2016_1,Z2016_2,SZ2018}):
\rev{\begin{eqnarray}
Z &=& \int D\bar{\psi}D\psi \, {\rm exp}\Big(  \int_{\cal M} \frac{d^D {p}}{|{\cal M}|} \nonumber\\&&\bar{\psi}^T({p}){Q}(p - A(i\partial_p))\psi({p}) \Big),\label{Z01}
\end{eqnarray}}
where the products of operators inside expression ${\cal Q}(p - A(i\partial_p))$ are symmetrized. Notice that the Peierls substitution $p \to p - A(i\partial_p))$ is precise. It is guaranteed by gauge invariance, and appears in the description of both relativistic and non - relativistic systems.

We relate operator $\hat{Q} = Q(p-A(i\partial_p))$ and its inverse $\hat{G} = \hat{Q}^{-1}$ defined in Hilbert space ${\cal H}$ of functions (on $\cal M$) with their matrix elements ${\cal Q}(p,q)$ and ${\cal G}(p,q)$ correspondingly:
$$
{\cal Q}(p,q) = \langle p|\hat{Q}| q\rangle, \quad {\cal G}(p,q) = \langle p|\hat{Q}^{-1}| q\rangle\,.
$$
Here the basis elements of $\cal H$ are normalized as $\langle p| q\rangle = \delta^{(D)}(p-q)$. Those operators obey the following equation
$$
\langle p|\hat{Q}\hat{G}|q\rangle = \delta({p} - {q})\,.
$$

Eq. (\ref{Z01}) may be rewritten as follows
\rev{\begin{eqnarray}
Z &=& \int D\bar{\psi}D\psi \, {\rm exp}\Big(  \int_{\cal M} \frac{d^D {p}_1}{\sqrt{|{\cal M}|}} \int_{\cal M} \frac{d^D {p}_2}{\sqrt{|{\cal M}|}}\nonumber\\&&\bar{\psi}^T({p}_1){\cal Q}(p_1,p_2)\psi({p}_2) \Big),\label{Z1}
\end{eqnarray}}
while the Green function of Bloch electron is given by
\rev{\begin{eqnarray}
{\cal G}_{ab}(k_2,k_1)&=& \frac{1}{Z}\int D\bar{\psi}D\psi \, {\rm exp}\Big(  \int_{\cal M} \frac{d^D {p}_1}{\sqrt{|{\cal M}|}} \int_{\cal M} \frac{d^D {p}_2}{\sqrt{|{\cal M}|}}\nonumber\\&&\bar{\psi}^T({p}_1){\cal Q}(p_1,p_2)\psi({p}_2) \Big) \frac{\bar{\psi}_b(k_2)}{\sqrt{|{\cal M}|}} \frac{\psi_a(k_1)}{\sqrt{|{\cal M}|}}\label{G1}\,.
\end{eqnarray}}
Here indices $a,b$ enumerate the components of the fermionic fields. In the following we will omit those indices for brevity.

\rvv{Notice, that we use the relativistic units, in which both $\hbar $ and $c$ are equal to unity.  Besides, elementary charge $e$ is included to the definition of electric and magnetic fields. }

\section{Wigner - Weyl formalism for the lattice models}

\label{Sect3}

The Wigner transformation of $\cal G$ is defined as the Weyl symbol of $\hat{G}$:
\begin{equation} \begin{aligned}
{G}_W(x,p) \equiv \int_{\cal M} dq e^{ix q} {\cal G}({p+q/2}, {p-q/2})\label{GWx}
\end{aligned}\,.
\end{equation}
\rvv{Correspondingly, the Weyl symbol of operator $\hat{Q}$ is given by ${Q}_W(x,p) \equiv \int_{\cal M} dq e^{ix q} {\cal Q}({p+q/2}, {p-q/2})$. It appears that for the slowly varying field $A(x)$ we have ${Q}_W(x,p) = {Q}_W(p-A(x))\equiv {Q}(p-A(x))$ (see \cite{Z2016_1}).}
\rv{It is assumed here that $Q(p_1,p_2)$ is nonzero for the values of $|p_1-p_2|$ much smaller than the size of the Brillouin zone. (The values of $|p_1 + p_2|$ may be arbitrary.) This occurs if the external electromagnetic field is slowly varying, i.e. its variation on the distance of the order of the interatomic distance may be neglected. Under these conditions the Wigner transformed Green function obeys the Groenewold equation (see \cite{Z2016_1,Z2016_2}) and Appendix H of the present paper}:
\begin{equation} \begin{aligned}
{G}_W(x_n,p) \star Q_W(x_n,p) = 1 \label{Geq}
\end{aligned} \,,
\end{equation}
that is
\begin{equation}\begin{aligned}
&1 =
{G}_W(x_n,p)
e^{\frac{i}{2} \left( \overleftarrow{\partial}_{x_n}\overrightarrow{\partial_p}-\overleftarrow{\partial_p}\overrightarrow{\partial}_{x_n}\right )}
Q_W(x_n,p)
\label{GQW}\end{aligned}\,.
\end{equation}
By $x_n$ we denote the lattice points. Although the lattice points are discrete, the differentiation over $x_n$ may be defined following \cite{Z2016_1,Z2016_2} because the functions of coordinates may be extended to their continuous values.

Let us define the (Grassmann - valued) Wigner function as
\rev{$$
W(p,q) = \frac{\bar{\psi}(p)}{\sqrt{|{\cal M}|}}\frac{{\psi}(q)}{\sqrt{|{\cal M}|}}\,.
$$}
We may also define the operator $\hat{W}[\psi,\bar{\psi}]$, whose matrix elements are equal to $W(p,q) = \langle p|\hat{W}[\psi,\bar{\psi}]|q\rangle$. The functional trace of an operator $\hat{U}$ is defined as
$$
{\bf Tr} \hat{U} \equiv \int dp {\rm Tr} \langle p | \hat{U}|p\rangle\,.
$$

Then the partition function receives the form:
\rev{\begin{eqnarray}
Z& =& \int D\bar{\psi}D\psi \, {\rm exp}\Big( {\bf Tr} \hat{W}[\psi,\bar{\psi}] \hat{Q} \Big)\,.\label{Z2}
\end{eqnarray}}
In the following by $W_W$ we denote the Weyl symbol of $\hat{W}$.

In general case the calculation of the Weyl symbol $Q_W$ of an operator $\hat{Q}=Q(p-A(i\partial_p))$ is a rather complicated problem. It has been solved in \cite{SZ2018} for the particular case of lattice Wilson fermions. The result is represented in Appendix A. It follows from those results that if the field $A$ is slowly varying, i.e. it almost does not vary on the distance of the order of the lattice spacing, we are able to substitute the sum over the lattice points by the integral, and take ${Q}_W(p,x)= Q_W(p-A(x)) \equiv Q(p-A(x))$. Of course, this refers not only to the lattice Wilson fermions, but to any lattice fermion model.

\rvv{It is worth mentioning, that there are several different definitions of the Wigner transformation for the lattice models. In this respect it is instructive to refer here to the definition proposed long time ago by F.Buot (see, e.g. \cite{buot,buot0,buot1}, another alternative definition may be found in \cite{kaspersky}).} \zz{However, in the present paper we use the different definition of the Weyl symbol of an operator $\hat A$}:
{\begin{equation} \begin{aligned}
{A}^{}_W(x,p) \equiv \int_{\cal M} dq e^{ix q} \langle {p+q/2}| \hat{A} | {p-q/2}\rangle\label{GWx}
\end{aligned}\,,
\end{equation}
\zz{where the integration is over momentum space $q\in {\cal M}$ (the Buot's definition being extended to the inhomogeneous systems considered here would give the definition of Eq. (\ref{GWx}), in which the integration is extended to the values of $q$, that obey $q/2 \in {\cal M}$).
It is worth mentioning that our definition of Eq. (\ref{GWx})} does not allow to obtain certain exact relations typical for the Buot's version of lattice Wigner - Weyl calculus. However, if the operator $\hat A$ is almost diagonal, then several useful identities follow. In particular, we have the star identity derived in Appendix H:
$$
(AB)_W = A_W\star B_W
$$
Besides,
$$
{\bf Tr} \hat{A}\hat{B} = \int d^Dx\,\frac{d^Dp}{(2\pi)^D}\,{\rm Tr} \, A_W(x,p) \star B_W(x,p)
$$
If, in addition, the periodical boundary conditions in $x$ - space are used, then  also
$$
{\bf Tr} \hat{A}\hat{B} = \int d^Dx\,\frac{d^Dp}{(2\pi)^D}\,{\rm Tr} \, A_W(x,p)  B_W(x,p)
$$
(We take the lattice with large linear size $L$ and periodic boundary conditions, at the end of calculations $L\to \infty$ as usual in many realizations of quantum field theory.)
Notice, that for the case, when external electric field is present, however,
$$
\int d^Dx\,\frac{d^Dp}{(2\pi)^D}\,{\rm Tr} \, A_W(x,p) \star B_W(x,p) \ne \int d^Dx\,\frac{d^Dp}{(2\pi)^D}\,{\rm Tr} \, A_W(x,p)  B_W(x,p)
$$
As it was mentioned above, the operator $\hat{Q} = Q(\hat{p}-A(\hat{x}))$ is almost diagonal (i.e. its matrix elements $\langle p|\hat{Q}| q\rangle$ do not vanish only for $|p-q|\ll 2\pi$) if the field $A$ varies slowly at the distances of the lattice spacing. Such fields correspond to the magnitudes of magnetic fields much smaller than thousands Tesla and wavelengths much larger than several Angstroms.  }\zz{These conditions allows us to obtain the above identities with the integrals over $x$ rather than the sums over the discrete lattice points (for the details see Appendix H).}

\section{Electric current}

\label{Sect4}

Variation of partition function gives the following expression for the \rev{average total current, i.e. the time average of the integral over the whole space of the electric current density} (the derivation is given in Appendix B):
\rev{\begin{eqnarray}
\langle J^k \rangle &=& - T\, \int d^D x \, \int \frac{d^D p}{(2\pi)^D}\, {\rm Tr} \, G_W(p,x)  \partial_{p_k} Q_W(p - { A}(x)). \label{J3}
\end{eqnarray}}
\rvv{Here $T$ is temperature. Since we are interested mainly in the theory at vanishing temperature, it is supposed to be set to zero at the end of calculations using Eq. (\ref{J3}).}
In the presence of periodic (in space) boundary conditions this expression is the topological invariant, i.e. the total current is not changed when the system is modified continuously.

\rvv{The appearance of the Hall current is a kinetic out of equilibrium phenomenon. However, in linear response theory it may be considered using the vacuum averages within the equilibrium theory. This is the content of the seminal fluctuation - dissipation theorem. Very briefly, we consider the response of electric current to the electric field as if the theory is in equilibrium. The answer is written as conductivity times electric field. The expression for the conductivity is to be calculated within the equilibrium theory, where the electric field is absent. This procedure is {totally equivalent} to the derivation of the Kubo formula for conductivity, that allows to calculate the conductivity (which is, of course, a quantity typical for the kinetic theory) using the {\it equilibrium} functional integral over fields. Technically, in the equilibrium theory the term bilinear in electromagnetic potential has the form
$${\rm log} \, Z^{(2)} = -\frac{1}{2} \int d^D x d^D y \, \Pi^{\mu\nu}(x,y) A_\mu(x) A_\nu(y),$$ where $\Pi^{\mu\nu}$ is the polarization tensor. The electric current given by Eq. (\ref{J3}) may then be represented as
\begin{equation}
\langle J^k \rangle =  T\, \int d^D x d^D y \Pi^{k\nu}(x,y) A_\nu(y)\label{Jk}
\end{equation}
It may be shown using the quantum kinetic theory \cite{LL10} that although Eq. (\ref{Jk}) is obtained within the equilibrium theory, for the purpose of the calculation of conductivity we are able to substitute into Eq. (\ref{Jk}) the electric potential that corresponds to the constant external electric field $E_k$. In this way one comes to the Kubo expression for the conductivity. (Wick rotation has to be taken into account here. Namely, we identify $A_{Dk} = \partial_D A_k-\partial_k A_D$ with $-i E_k$.) The given algorithm will be used below to derive the expression for the Hall conductivity within the Wigner - Weyl formalism. Namely, we will take formally the equilibrium expression for electric current of Eq. (\ref{J3}), calculate its response to external field strength, substitute the external field strength by the expression corresponding to external electric field, and consider the resulting expression for the response as the Hall conductivity.  }

In Appendix C we remind the reader the consideration of the anomalous QHE and represent the results obtained in \cite{Z2016_1,Z2016_2}. It appears that the response of the electric current to weak external field strength $A_{ij} = \partial_i A_j - \partial_j A_i$ is the topological invariant in momentum space. This invariant is composed of the one - particle Green function (that is the function of the only momentum $p$ for $A=0$). In the present paper we propose the generalization of this representation to the case, when in addition to the weak external field strength $A_{ij}$ there is the strong external field strength $B_{ij} =  \partial_i B_j - \partial_j B_i$. The former gives rise to the external electric field while the latter corresponds to strong essentially varying external magnetic field that provides the system with the gaps and the topological structure responsible for the QHE. In this situation the one - particle Green function at $A=0$ depends on two momenta because of the presence of the field $B(x)$. So, we discuss the case, when the fermions are in the presence of two Abelian gauge fields $A_{i}$ and $B_i$.

Let us first consider the case of constant external field strength $A_{ij}$ and arbitrary field $B(x)$. We are going to expand $\langle J \rangle$ in powers of $A_{ij}$ and to keep the linear term only. At the same time the field $B(x)$ is taken into account completely. We start from
\rev{\begin{eqnarray}
\langle J^k \rangle &=&  - T\, \int d^D x \, \int \frac{d^D p}{(2\pi)^D}\, {\rm Tr} \, G_W(p,x) \nonumber\\&&  \partial_{p_k} Q_W(p - { A}(x)- B(x)) \Big)
\label{J30}
\end{eqnarray}}
and define $G_W$ and $G^{(0)}_W$ that obey
$$
G_W(p,x) \star Q_W(p - A(x)- B(x)) =1
$$
and
$$
G^{(0)}_W(p,x) \star Q_W(p - B(x))=1\,.
$$
We also denote $Q_W(p,x) = Q_W(p - A(x)- B(x))$. The Wigner transformed inverse propagator in the presence of external magnetic field (without external electric field) is $Q^{(0)}_W(p,x)=Q_W(p - B(x))$.
Let us represent $Q_W$ as
$$
Q_W(p,x) \approx  Q^{(0)}_W(p,x) -  \partial_{p_m} Q^{(0)}_W(p,x) A_m(x)
$$
and
 $G_W$ as follows
\begin{eqnarray}
G_W(p,x) &\approx & G^{(0)}_W(p,x) + G^{(0)}_W(p,x) \nonumber\\&&\star \partial_{p_m} Q^{(0)}_W(p,x) A_m(x) \star G^{(0)}_W(p,x)\nonumber \\
 & \equiv & G_W(p,x|y)\Big|_{y=x}\,.
\end{eqnarray}
Here
$$
G_W(p,x|y)= G^{(0)}_W(p,x) + G^{(0)}_W(p,x) $$
$$\star \partial_{p_m} Q^{(0)}_W(p,x) A_m(y) \star G^{(0)}_W(p,x)\,,
$$
and
$$
\star = \circ \, \ast
$$
where
$
\ast = e^{\frac{i}{2} \left( \overleftarrow{\partial}_{x}\overrightarrow{\partial_p}-\overleftarrow{\partial_p}\overrightarrow{\partial}_{x}\right )}
$
while
$
\circ = e^{\frac{i}{2} \left( \overleftarrow{\partial}_{y}\overrightarrow{\partial_p}-\overleftarrow{\partial_p}\overrightarrow{\partial}_{y}\right )}\,.
$
The direct check shows that in Eq. (\ref{J3}) the terms proportional to $A$ (that do not contain the derivatives of $A$)  cancel each other, so that the first relevant terms are proportional to $\partial_i A_j$. We also may represent Eq. (\ref{J3}) as
\rev{\begin{eqnarray}
\langle J^k \rangle &=&  - T\, \int d^D x \, \int \frac{d^D p}{(2\pi)^D}\, {\rm Tr} \Big(\, G_W(p,x|y) \nonumber\\&&\circ \ast \partial_{p_k} Q_W(p - { A}(y)- B(x)) \Big)_{y=x}\,.
\label{J31}
\end{eqnarray}}
$G_W(p,x) = G_W(p,x|x)$ obeys equation
 $$
G_W(p,x|y) \circ \, \ast Q_W(p - A(y)- B(x))\Big|_{y=x} =1\,.
$$
Up to the terms linear in the derivatives of $A$ we get:
$$
\partial_{y_k} G_W(p,x|y) = \partial_{y_k} G^{(0)}_W(p,x) \star \partial_{p_m} Q^{(0)}_W(p,x)\,, $$
$$A_m(y) \star G^{(0)}_W(p,x) \approx G^{(0)}_W(p,x) \ast \partial_{p_m} Q^{(0)}_W(p,x)$$$$ \partial_{y_k} A_m(y) \ast G^{(0)}_W(p,x)
$$
$$
= - \partial_{p_m}G^{(0)}_W(p,x) \partial_{y_k} A_m(y)\,.
$$
This allows to rewrite
 $$
G_W(p,x|y) \, e^{-\frac{i}{2} \overleftarrow{\partial}_{p_j}\overrightarrow{\partial_{p_i}}A_{ij}}\, \ast Q_W(p - A(y)- B(x))\Big|_{y=x} =1\,.
$$
Up to the terms linear in $A_{ij}$ this equation receives the form
$$
G_W(p,x|y) \, \Big(1-\frac{i}{2} \overleftarrow{\partial}_{p_j}\overrightarrow{\partial_{p_i}}A_{ij}\Big)\, \ast Q_W(p - A(y)- B(x))\Big|_{y=x} =1\,.
$$
Its solution is
\begin{eqnarray}
 G_W(p,x|x)  &= & G_W^{(0)}\nonumber\\&& + G^{(0)}_W\star \partial_{p_m} Q^{(0)}_W  \star G^{(0)}_W A_m \nonumber\\&&\cor{+} \frac{i}{2}  G_W^{(0)}\ast \frac{\partial  Q^{(0)}_W}{\partial p_i} \ast G_W^{(0)} \ast \frac{\partial  Q^{(0)}_W}{\partial p_j} \ast G_W^{(0)}
A_{ij}\,.
\end{eqnarray}
We rewrite Eq. (\ref{J3}) as follows:
\rev{\begin{eqnarray}
\langle J^k \rangle &=& {-} T\, \int d^D x \, \int \frac{d^D p}{(2\pi)^D}\, {\rm Tr} \Big(\, G_W(p,x|y) e^{-\frac{i}{2} \overleftarrow{\partial}_{p_j}\overrightarrow{\partial_{p_i}}A_{ij}}\nonumber\\ && \ast \partial_{p_k} Q_W(p - { A}(y)- B(x)) \Big)_{y=x}\,,
\label{J31}
\end{eqnarray}}
 and obtain:
\rev{\begin{eqnarray}
\langle J^k \rangle  &\approx & - T\, \int d^D x \, \int \frac{d^D p}{(2\pi)^D}\, {\rm Tr} \, G^{(0)}_W(p,x) \ast \partial_{p_k}  Q^{(0)}_W(p,x)  \nonumber\\ && \cor{-}  \frac{i T}{2}\, \int d^D x \, \int \frac{d^D p}{(2\pi)^D}\, {\rm Tr} \,  G^{(0)}_W\ast \frac{\partial  Q^{(0)}_W}{\partial p_i}\nonumber\\ && \ast G^{(0)}_W \ast \frac{\partial  Q^{(0)}_W}{\partial p_j} \ast G^{(0)}_W \ast \partial_{p_k} Q^{(0)}_W
A_{ij}\,.
\label{J32}
\end{eqnarray}}
\rv{The first term here is the equilibrium ground state current in the absence of electric field that according to the Bloch theorem vanishes (see \cite{Yamamoto} and references therein). It is worth mentioning that the general proof of this statement is absent in the framework of quantum field theory/condensed matter theory with relativistic spin - orbit interactions taken into account. We are aware of the clear proof of the Bloch theorem in the nonrelativistic quantum mechanics only. There may, in principle, be some marginal exclusions. However, we omit here consideration of such marginal cases and assume that the Bloch theorem in its conventional form is valid. In Appendix D we present the proof that the first term in Eq. (\ref{J32}) is the topological invariant, i.e. it is not changed when the system is modified smoothly. It complements the mentioned Bloch theorem and shows that this term remains constant (and is equal to zero according to the Bloch theorem) when the system is modified smoothly. Anyway, this term cannot contribute to the electric current of the QHE because it does not contain external electric field.  }

For $D=4$ the second term in this expression may be rewritten as
\begin{eqnarray}
\langle J^{k}\rangle  & \approx & \cor{-} \frac{\cal V}{4\pi^2}\epsilon^{ijkl} {\cal M}_{l} A_{ij}, \label{calM0}\,,\\
{\cal M}_l &=& \frac{1}{\int d^4 x}  \int_{} \,{\rm Tr}\, \nu_{l} \,d^4p d^4 x \nonumber \\
&=& \frac{T}{\cal V}  \int_{} \,{\rm Tr}\, \nu_{l} \,d^4p d^4 x \label{Ml} \,\nonumber \\
\nu_{l} & = &  \rev{-\frac{i}{3!\,8\pi^2}}\,\epsilon_{ijkl}\, \Big[  {G}_W \ast \frac{\partial {Q}_W}{\partial p_i}\ast \frac{\partial  {G}_W}{\partial p_j} \ast \frac{\partial  {Q}_W}{\partial p_k} \Big]_{A=0}\,, \nonumber
\end{eqnarray}
{where ${\cal V}$ is the spacial volumn of the system.}
It is worth mentioning, that in the present paper the star product of several functions $ f_1(p,x), f_2(p,x),...,f_n(p,x)$ should be understood as
$$
f_1 * f_2 * ... * f_n = [...[[f_1 * f_2] * f_3] * ... * f_n]\,.
$$
Here each operation $*$ acts within the nearest brackets only, that is the derivatives entering the star act only within those brackets. At the same time the property $[[f_1 * f_2] * f_3] = [f_1 * [f_2 * f_3]]$ {(the proof is given in Appendix H)} allows to move the brackets within the multiple products, and to write the product without brackets at all.
For the two - dimensional systems ($D=3$) we have:{
\begin{equation}
\langle J^{k} \rangle \approx \cor{-}\frac{\cal S}{4\pi}{\cal M} \epsilon^{ijk} A_{ij}\,,\label{JA2D0}
\end{equation}}
{where ${\cal S}$ is the area of the system.}
Here  ${\cal M}$ is the topological invariant in phase space
\begin{eqnarray}
{\cal M} &= &  \frac{1}{\int d^3 x} \int \,{\rm Tr}\, \nu_{} \,d^3p d^3x \nonumber\\
&= &  \frac{T}{\cal S} \int \,{\rm Tr}\, \nu_{} \,d^3p d^3x\, ,\nonumber
\\
\nu_{} & = &    \rev{-\frac{i}{3!\,4\pi^2}}\,\epsilon_{ijk}\, \Big[  {G}_W(p,x)\ast \frac{\partial {Q}_W(p,x)}{\partial p_i} \,,\nonumber\\ &&\ast \frac{\partial  {G}_W( p,x)}{\partial p_j} \ast \frac{\partial  {Q}_W(p,x)}{\partial p_k} \Big]_{A=0}  \label{calM2d2}\,.
\end{eqnarray}
The proof that Eq. (\ref{calM2d2}) is indeed the topological invariant is given in Appendix D. In the similar way it may be proved that Eq. (\ref{calM0}) represents the topological invariant. \rv{As it was mentioned above, for the validity of Eqs. (\ref{calM2d2}) and (\ref{calM0}) we need slowly varying potentials $A(x)$. This gives the two conditions: the magnitude of the external magnetic field is much smaller than thousands of Tesla, and the wavelength is much larger than the interatomic distance.}

\rvv{Eqs. (\ref{calM0}) and (\ref{calM2d2}) consist the main result of our paper. It is worth mentioning that although we derived those expressions for lattice models, they remain valid for the continuous models as well. In order to obtain the expression for Hall current we substitute to Eqs. (\ref{calM0}) and (\ref{calM2d2}) the Euclidean field strength $A_{ij}$ corresponding to the electric field $E_k$. Its nonzero components are \begin{equation}
A_{Dk} = \partial_D A_k-\partial_k A_D = -i E_k.
\end{equation}
This results in the following expression for the current density integrated over the whole volume divided by this volume and averaged with respect to time) in the presence of the external electric field $E_i$:
\begin{eqnarray}
\langle j^{k}\rangle   = \cor{-} \frac{1}{2\pi^2}\epsilon^{kjl4} {\cal N}_{l} E_{j} \label{j03},\, {\cal N}_l =  - \frac{T\epsilon_{ijkl}}{{\cal V}\, 3!\,8\pi^2}\ \int d^4 x d^4p \,{\rm Tr}\,  \Big[  {G}_W(p,x) \ast \frac{\partial {Q}_W(p,x)} {\partial p_i}\ast \frac{\partial  {G}_W(p,x)}{\partial p_j} \ast \frac{\partial  {Q}_W(p,x)}{\partial p_k} \Big] \end{eqnarray}
for the $3+1$D system (here $\cal V$ is the overall volume), and
\begin{eqnarray}
\langle j^{k} \rangle =  \cor{-}\frac{1}{2\pi} {\cal N} \epsilon^{3kj} E_{j}\label{j02},\,
{\cal N} =    \frac{T\epsilon_{ijk}}{{\cal S}\,3!\,4\pi^2} \int d^3p  d^3x     \, {\rm Tr}  \, \Big[  {G}_W(p,x)\ast \frac{\partial {Q}_W(p,x)}{\partial p_i} \ast \frac{\partial  {G}_W( p,x)}{\partial p_j} \ast \frac{\partial  {Q}_W(p,x)}{\partial p_k} \Big]
\end{eqnarray}
for the $2+1$ D system (here $\cal S$ is the area of the system) }

\rvv{It is worth mentioning, that for the gapless systems when the two - point Green function has poles, the singularities may appear in the integrals in Eqs. (\ref{j03}) and (\ref{j02}). As a result those expressions for the conductivities may become not well defined. In the absence of interactions this occurs if chemical potential crosses one (or more) of the energy branches. We come to the conclusion, that for the gapped system the given above expressions for $\cal N$ (${\cal N}_l$) are well defined while for the gapless systems there may occur the uncertainty in their definition. Even if those expressions remain well defined, they are already not the topological invariants for the gapless systems because the appearance of the singularities breaks the proof presented in Appendix D (the examples are given by the so - called Weyl semimetals \cite{Z2016_1}). For the case of constant external magnetic field the system is gapped when chemical potential remains between the Landau Levels, and it becomes gapless if it coincides with one of them. Obviously, weak variations of magnetic field (as well as a weak disorder) does not disturb this pattern: the energy branches become thick, but the system remains gapped if the chemical potential does not cross them. Sufficiently strong variations of magnetic field break this pattern. Therefore, the above Eqs. (\ref{j03}), (\ref{j02}) are to be taken with care in this case. They still remain relevant for the gapped system. The consideration of the precise conditions needed for the system to be gapped in the presence of strongly varying magnetic field remains out of the scope of the present paper. }

\section{The alternative derivation of Hall conductivity for the case of constant magnetic field}

\label{Sect5}

Notice, that Eq. (\ref{calM2d2}) may be reduced to the TKNN invariant for the case of the electrons in crystals without interactions and with $Q(p) = i\omega  - {\cal H}(p_x,p_y)$, where $\cal H$ is the one - particle Hamiltonian. This occurs if either there is no external magnetic field or if it is  constant.
The case of continuum model (say, with ${\cal H}(p_x,p_y) = \frac{p_x^2+p_y^2}{2M}$) in the presence of constant magnetic field ${\cal B}$ (orthogonal to the plane of the given $2D$ system) is the simplest case, when Eq. (\ref{calM2d2}) is not reduced to the TKNN invariant. Below we illustrate our result by the consideration of such a system.

We choose the gauge, in which the gauge potential is
$$
B_x = 0, \quad B_y = {\cal B} x\,.
$$
External electric field ${\it E}_y$ corresponding to the gauge potential $A$ is directed along the axis $y$. The above derived expressions give the following expression for the electric current averaged over the area of the system:{
\begin{equation}
\langle j_{x} \rangle \approx \cor{-}\frac{\cal S}{2\pi}{\cal N} {\it E}_y\,.
\end{equation}}
Here the average Hall conductivity is given by \rev{$\sigma_{xy} = -{\cal N}/2 \pi$ (recall that $j_x = \cor{+}\sigma_{xy} E_y$)} while ${\cal N}=i{\cal M}$ is the following topological invariant in phase space
\begin{eqnarray}
{\cal N}& = & {\frac{T}{3! 4\pi^2 {\cal S}}} \, \int \,{\rm Tr}\, d^3p \, d^3x\,   \epsilon_{ijk}\, \Big[  {G}_W(p,x)\ast \frac{\partial {Q}_W(p,x)}{\partial p_i} \nonumber\\&& \ast \frac{\partial  {G}_W( p,x)}{\partial p_j} \ast \frac{\partial  {Q}_W(p,x)}{\partial p_k} \Big]_{A=0}\,.
  \label{calM2d23_}
\end{eqnarray}

For the practical calculations in the given particular case it is more useful to represent this expression in terms of the Green function written in momentum representation:
\begin{eqnarray}
{\cal N} &=& {\frac{T(2\pi)^3}{3!\,4\pi^2{\cal S}}}\, \int \,{\rm Tr}\, d^3p^{(1)} \, d^3p^{(2)}\, d^3 p^{(3)} \, d^3 p^{(4)}\,       \epsilon_{ijk}\,\nonumber\\&& \Big[  {G}(p^{(1)},p^{(2)})\Big( [\partial_{p^{(2)}_i} + \partial_{p^{(3)}_i}] Q(p^{(2)},p^{(3)})\Big)\nonumber\\&&  \Big( [\partial_{p^{(3)}_j} + \partial_{p^{(4)}_j}]  G(p^{(3)},p^{(4)}) \Big) \nonumber\\&&\Big( [\partial_{p^{(4)}_k} + \partial_{p^{(1)}_k}] Q(p^{(4)},p^{(1)})\Big) \Big]_{A=0}\,.
  \label{calM2d23P_}
\end{eqnarray}
This representation has been deduced from Eq. (\ref{calM2d23_}) in  Appendix E. There also the transformation of Eq. (\ref{calM2d23P_}) to the following form has been derived:
{\begin{eqnarray}
{\cal N} &=&   -\frac{2\pi i}{\cal S}\,\sum_{n,k}   \, \epsilon_{ij}\,\Big[  \frac{ \langle n| [{\cal H}, {\hat x}_i] | k \rangle    \langle k | [{\cal H}, {\hat x}_j] | n \rangle }{({\cal E}_k - {\cal E}_n)^2} \Big]_{A=0}\nonumber\\&&\theta(-{\cal E}_n)\theta({\cal E}_k)\label{sigmaHH_}\,.
\end{eqnarray}}
By operator $\hat x$ we understand $i\partial_{p}$ acting on the the wavefunction written in momentum representation:
$$
\hat{x}_i \Psi(p) = \langle p|\hat{x}_i |\Psi\rangle = i\partial_p \langle p|\Psi\rangle = i \partial_p \Psi(p)\,.
$$
Eq. (\ref{sigmaHH_}) (divided by $2\pi$) is the standard expression for the Hall conductance that follows from Kubo formula. For completeness we present the derivation of Eq. (\ref{sigmaHH_})  from the Kubo formula in Appendix F (see also \cite{Tong:2016kpv,Landsteiner:2012kd}). The further calculation of Hall conductance using Eq. (\ref{sigmaHH_}) is also standard (see, for example, \cite{QHEB}). We give it in Appendix G. The result reads
{\begin{eqnarray}
{\cal N}
&=&  N \, {\rm sign}(-{\cal B})\,,
  \label{calM2d233}
\end{eqnarray}}
while $N$ is the number of the occupied branches of spectrum. This way we came to the conventional expression for the Hall resistivity of a  system of electrons with charge $-|e|$ in the presence of constant magnetic field ${\cal B} = - |e| B_z$ (directed along the $z$ axis) and constant electric field:
\rev{
$$
\rho_{xy} = \frac{e}{\sigma_{xy}} = \frac{2\pi \hbar }{e^2 N}\,{\rm sign}(B_z)\,.
$$
Here we restore the conventional units with $\hbar \ne 1$. Recall, that the resistivity tensor is related to the conductivity tensor as follows:
$$
\left(\begin{array}{cc} 0 & \sigma_{xy} \\ \cor{-}\sigma_{xy} & 0\end{array}\right) = \left(\begin{array}{cc} 0 & \cor{-}\rho_{xy} \\   \rho_{xy} & 0\end{array}\right)^{-1}\,.
$$
}

\section{Conclusions and discussion}

\label{SectConcl}

\rev{In the present paper we derive the representation of Hall conductivity as the topological invariant in phase space composed of the Wigner transformed Green functions. Our main result is given by the following expressions for the average Hall current (i.e. the current density integrated over the whole volume divided by this volume and averaged with respect to time}) in the presence of the external electric field $E_i$:
\rev{\begin{eqnarray}
\langle j^{k}\rangle   = \cor{-}  \frac{1}{2\pi^2}\epsilon^{kjl4} {\cal N}_{l} E_{j} \label{calM0C},\, {\cal N}_l =  - \frac{T\epsilon_{ijkl}}{{\cal V}\, 3!\,8\pi^2}\ \int d^4 x d^4p \,{\rm Tr}\,  \Big[  {G}_W(p,x) \ast \frac{\partial {Q}_W(p,x)} {\partial p_i}\ast \frac{\partial  {G}_W(p,x)}{\partial p_j} \ast \frac{\partial  {Q}_W(p,x)}{\partial p_k}  \Big]\label{rez1} \end{eqnarray}
for the $3+1$D system (here $\cal V$ is the overall volume)}, and
\rev{\begin{eqnarray}
\langle j^{k} \rangle =  \cor{-}\frac{1}{2\pi} {\cal N} \epsilon^{3kj} E_{j}\label{JA2D0C},\,
{\cal N} =    \frac{T\epsilon_{ijk}}{{\cal S}\,3!\,4\pi^2} \int d^3p  d^3x     \, {\rm Tr}  \, \Big[  {G}_W(p,x)\ast \frac{\partial {Q}_W(p,x)}{\partial p_i} \ast \frac{\partial  {G}_W( p,x)}{\partial p_j} \ast \frac{\partial  {Q}_W(p,x)}{\partial p_k} \Big]\label{rez2}
\end{eqnarray}
for the $2+1$ D system (here $\cal S$ is the area of the system)}.
Those expressions may be used both for the description of the intrinsic AQHE and for the description of the QHE in the presence of \zz{varying} magnetic field. \zz{Notice, that the same approach may be applied to the inhomogeneous systems, in which the inhomogeneity has another origin. For example, in \cite{Fialkovsky:2019nso} the Hall conductivity has been considered for the systems in the presence of elastic deformations.}

\zz{It is the common lore, that the QHE conductivity is given by the counting of the occupied gapped energy levels, while the role of the magnetic field is to provide that the spectrum is gapped, and to provide each energy level with the nontrivial topology. For the case of constant magnetic field the value of Hall conductivity of an ideal system is given below in Appendix G by the last line of Eq. (\ref{QHEB}). In this expression the Hall conductivity is proportional to the number of occupied states divided by the value $\cal B$ of magnetic field. For the Landau Levels (LL), when the degeneracy is proportional to magnetic field, $\cal B$ is cancelled, and we arrive at the Hall conductivity proportional to the number of the occupied LL. This result has been derived for sufficiently weak magnetic fields\footnote{This requirement corresponds to $|{\cal B}|\ll 10 000$ Tesla. Otherwise the expression for the Chern number carried by the LL is more nontrivial.}, and when magnetic field is constant. The fact that this result does not contain the value of $\cal B$ explicitly prompts that the weak variations of the latter should not affect the value of Hall conductance. However, to the best of our knowledge, this statement has never been proven explicitly earlier. We prove it in our present paper. Namely, we consider from the very beginning varying magnetic fields. We represent the Hall conductivity in Eqs. (\ref{rez1}), (\ref{rez2}) as the topological invariant written in terms of the Green function. This expression from the very beginning is valid for varying magnetic fields. Moreover, it remains valid in the presence of interactions as well \cite{ZZ2019_2}. The small variations of magnetic field cannot affect the value of this topological invariant (by the definition of the topological invariant). Now indeed if we start from the case of constant magnetic field, and turn on the small variations of $\cal B$, then the Hall conductivity is not changed if the Fermi energy remains between the energy levels (i.e. the Green function does not acquire the poles while turning on the variations). }

\zz{In the presence of constant magnetic field the energy levels are degenerate due to the specific symmetry. (The degeneracy is proportional to the value of magnetic field.) This degeneracy is lifted obviously if the symmetry is removed, i.e. if the magnetic field receives small variations. We refer, as an example, to \cite{varB} and references therein. To be explicit, in \cite{varB} the simple configuration of the inhomogeneous magnetic field in graphene has been considered, in which magnetic field is constant everywhere except for the region $|x|< d$, where it has the different value. The resulting branches of energy are represented in Fig. 2.a of \cite{varB} for the particular choice of parameters. One can see, that the Landau levels are modified (certain dependence on $k$ appears), but the system still remains gapped.  This illustrates the general case, when for small variations the change to the pattern of the energy levels is just the thickening of the Landau Levels. The more detailed analysis of the modification of the energy levels remains out of the scope of the present paper, and is to be the subject of a separate investigation. The presented expressions may be used later in more practical calculations. To be explicit, one has to substitute to Eqs. (\ref{rez1}) or (\ref{rez2}) the particular form of the magnetic field depending on coordinates.  Suppose, that for a certain profile of the varying magnetic field there is a gap, the Fermi energy belongs to this gap\footnote{Under these conditions Eqs. (\ref{rez1}), (\ref{rez2}) are convergent.}, and the Hall conductivity is known. Then the topological nature of Eqs. (\ref{rez1}) or (\ref{rez2}) means that small modification of the dependence of the magnetic field on coordinates cannot change the value of Hall conductivity. } It has been mentioned in the Introduction that disorder pushes the Hall current towards the boundary. It could be that variations of magnetic field may do the same. But this remains out of the scope of the present paper.

\zz{The present paper actually is devoted first of all to the consideration of the case when the external magnetic field is varying. This variation may assume the presence of local electric currents. However, this does not occur necessarily. Let us fix the gauge $\nabla {\bf A}=0$. Then the local electric current corresponding to varying magnetic field is proportional to the Laplacian of electromagnetic potential $\bf A$. In the gapped systems without external electric field, and without impurities the macroscopic electric currents do not appear if Fermi energy remains within the gap. We come to the Laplace equation $\Delta {\bf A} = 0$. It may have nontrivial solution for certain boundary conditions giving rise to varying magnetic field. The different situation appears if we require that at infinity the electromagnetic potential remains the same as the one that leads to the constant magnetic field. Then in the absence of local electric current within the bulk the solution of Laplace equation does not admit the variations of magnetic field. Therefore, if we want to modify the constant magnetic field in such a way, that $\bf A$ remains the same at infinity while within the limited region of the bulk the magnetic field varies, then the local electric currents in the bulk are inevitable. Therefore, speaking of weak modification of constant magnetic field we are forced to assume the appearance of the local electric currents within the bulk. In the presence of impurities their electric field gives rise to various local currents, which, in turn, influence magnetic field and lead to its variations. However, if we want to consider the ideal crystals, then the gap does not allow the appearance of the local currents, and we have to assume the appearance of the impregnations of the other types of matter, which carry various currents (the integral of electric current over the system remains zero). This is the way to introduce varying magnetic field to the idealized pure system. }

 \rvv{Weak disorder may be introduced via the modification of electromagnetic potential $B_\mu$ entering Eq. (\ref{J30}). Above we assumed, that this field gives rise to varying magnetic field. However, we may also suppose, that the component $B_D$ models the electric potential of impurities. If it varies slowly at the distance of the order of the lattice spacing, then our derivation of Eqs. (\ref{j03}), (\ref{j02}) remains valid. We come to the conclusion that Eqs. (\ref{calM0C}), (\ref{JA2D0C}) represent the average conductivity and are robust to the introduction of weak disorder.  The role of interactions is discussed in \cite{ZZ2019_2}, where it has been shown to all orders in fine structure constant, that Coulomb interactions between electrons do not affect the above proposed expressions of  Eqs. (\ref{calM0C}), (\ref{JA2D0C}). Consideration of the other interactions like the phonon exchange may be performed in the similar way. However, this remains out of the scope of the present paper. We refer here to the previous studies of the role of interactions in the Quantum Hall Effect: \cite{nocorrectionstoQHE,Altshuler0,Altshuler,AQHE_no_corr,corr_WSM1,corr_WSM2,2DTI_corr,Tang2018_Science}.
The role of interactions in the intrinsic Anomalous Quantum Hall effect has been discussed in \cite{ZZ2019}.}

{The mathematical structure of Eqs. (\ref{calM0C}), (\ref{JA2D0C}) repeats that of the topological invariants discussed in \cite{Volovik2003} with the ordinary product of matrices substituted by the star (Moyal) product of Wigner - Weyl formalism.}
The main advantage of the obtained expressions for the average conductivity is that they are written in terms of the Green functions. \rvv{It follows from the above discussion and from the results of \cite{ZZ2019_2} that weak disorder does not affect Eqs. (\ref{calM0C}), (\ref{JA2D0C}) while the only effect of (weak) interactions is the modification of the two - point Green functions entering those expressions. We should substitute there the ones of the interacting system. For the similar discussion of the homogeneous systems (see \cite{Matsuyama:1986us,Volovik0,Volovik2003,Z2016_1}).} Here it is extended to the inhomogeneous system in the presence of the arbitrarily varying magnetic field. Our formulas allow to show, in particular, that the average Hall conductivity is not changed when this magnetic field is modified smoothly (unless the topological phase transition is encountered). Therefore, the calculation of Hall conductance for the complicated configuration of magnetic field may be reduced to that of the more simple magnetic field profile.

\zz{We restrict ourselves by the magnitudes of the magnetic field much smaller than thousands Tesla. We also assume, that the typical distances at which the magnetic field may be changed essentially are much larger than the interatomic distance. Under these conditions the used formalism remains valid. One more condition is that the system remains gapped, and the Fermi energy belongs to the gap. Under this latter condition Eqs. (\ref{rez1}) and (\ref{rez2}) are convergent and are the topological invariants. Weaker conditions (when the Fermi energy does not necessarily belong to the gap) are not always inconsistent with the convergence of the mentioned expressions. However, then these expressions are not the topological invariants (the derivation of Appendix D is not valid if there is a pole of the Green function). Apart from the mentioned here conditions the variation of magnetic field remains arbitrary. As it was mentioned above, then the Hall conductivity is robust to smooth modification of the system. In particular, we may start from the case of the constant magnetic field, next modify the magnetic field slightly. The Hall conductivity remains the same until the phase transition is encountered. This occurs when strength of the variation becomes sufficiently large, the pole appears in the Green function, and the integrals in the expressions for the Hall conductivity become divergent.  We may start, however, from a certain profile of the magnetic field that is not constant, and varies strongly in space. Then the Hall conductivity is still given by  Eq. (\ref{rez1}) or (\ref{rez2}) if there is a gap, and Fermi energy belongs to it. Again, the smooth modification of magnetic field does not lead to the change of the Hall conductivity until a phase transition. }

 \rv{It is worth mentioning, that the Coulomb interactions in the case of sufficiently pure systems give rise to the fractional version of the QHE, which is out of the scope of the present paper. However, we expect, that the Wigner - Weyl formalism may be relevant for its description as well (see, e.g. \cite{Susskind,noncommutative}).}

The approach of \cite{Z2016_1} has been applied also for the investigation of the other non - dissipative currents (for the review see \cite{ZKA2018,ZK2018}). In this way the Chiral Separation Effect \cite{KZ2017}, the Chiral Vortical Effect \cite{AKZ2018}, and the Chiral Torsional Effect \cite{KZ2018} were related to the topological invariants. The methodology of the present paper may also be extended to those non - dissipative transport effects for the non - homogeneous systems, i.e. the systems with the arbitrarily varying magnetic field, rotation velocity, emergent torsion correspondingly. Finally, we would like to mention that another formulations of Wigner - Weyl formalism in lattice models have been proposed by various authors (see, for example, \cite{buot,buot0,buot1,kaspersky} and references therein).

The authors are grateful to M.Suleymanov and Chunxu Zhang for useful discussions, and to I.Fialkovsky for careful reading of the manuscript and useful comments.

\section*{Appendix A. Weyl symbol of Wilson Dirac operator}

\rv{In \cite{SZ2018} the Weyl symbol of  $\hat{Q} = Q(p-A(i\partial_p))$ \rev{(for the field $A(x)$ that varies slowly, i.e. if its variation at the distance of the order of lattice spacing is negligible)}  was calculated for the case of the so - called Wilson fermions (in four space - time dimensions) with
\begin{equation} \begin{aligned}
\hat{Q}({ p}) =\sum_{k=1,2,3,4} \gamma^k g_k ({ p})-im({ p})
\label{WF}\end{aligned} \end{equation}
where $\gamma^k$ are Euclidean $4\times 4$ Gamma - matrices. The corresponding spinors are four - component. Functions $g_k$ and $m$ are given by
\begin{equation} \begin{aligned}
g_k({ p})=\sin( p_k) \quad\quad m({ p})=
m^{(0)}+\sum_{\nu=1}^4 (1-\cos(p_\nu))\,.
\label{Z}\end{aligned} \end{equation}
We obtain
\begin{equation} \begin{aligned}
\Big[{Q}({p}-A(i{\partial_{ p}}))\Big]_W =\sum_{k=1,2,3,4} \gamma^k {\rm sin} ( p_k - {\cal A}_k({\bf x}))\\-i (m^{(0)}+\sum_{\nu=1}^4 (1-\cos(p_\nu - {\cal A}_{\nu}({\bf x}))))
\label{WF}\end{aligned} \end{equation}
where $\cal A$ is the following transformation of electromagnetic field:
\begin{equation}
{\cal A}_\mu({ x}) =  \int \big[\frac{\sin(k_{\mu}/2)}{k_{\mu}/2}\tilde A_\mu({ k})e^{i{ kx}}+c.c. \big]dk \label{calA}
\end{equation}
while the original electromagnetic field had the form:
$${A}_\mu({\bf x}) =  \int \big[\tilde A_\mu({k})e^{i{kx}}+c.c. \big]dk\,. $$
that is
$$
\tilde{A}(p) =  \frac{1}{|{\cal M}|}\sum_{x_n }  { A}(x_n)e^{-i{ k x_n}}
$$
In coordinate space we have
\begin{eqnarray}
{\cal A}_j(x) &=&  \int \big[\frac{\sin(k_j/2)}{k_j/2}\tilde A_j(k)e^{i { kx}}\big]dk\nonumber\\
&=&  \frac{1}{|{\cal M}|}\sum_{y_n} \int \big[\frac{\sin(k_j/2)}{k_j/2}{ A}_{j}(y_n)e^{-i{ky_n}}e^{i {kx}}\big]dk
\nonumber\\
&=& \sum_{y_n}{ A}_{j}(y_n) \int_{-1/2}^{1/2}dz \int \big[e^{i (k_{j}z + kx - ky_n)} \big] \frac{dk}{|{\cal M}|}\nonumber\\
&=& \sum_{y_n}{ A}_{j}(y_n) {\cal D}(x-y_n)
\end{eqnarray}
where
$$
{\cal D}(x) = \int_{-1/2}^{1/2}dz \int \big[e^{i( k_j z + kx)} \big] \frac{dk}{|{\cal M}|}
$$
One may check via the direct substitution, that
\begin{eqnarray}
{\cal A}_j(x) &=&  \int^{1/2}_{-1/2} {A}_j(x + e^{(j)}u) du
\end{eqnarray}
where $e^{(j)}$ is the unity vector in the $j$ - th direction:
\begin{eqnarray}
{\cal A}_j(p) & = & \frac{1}{|{\cal M}|}\sum_{x_n} {\cal A}_j(x_n) e^{-ip x_n}\nonumber \\ &=&  \frac{1}{|{\cal M}|}\int^{1/2}_{-1/2}du \sum_{x_n} { A}_{j}(x_n + e^{(j)}u)e^{-i p x_n}  \nonumber\\ &=&  \int^{1/2}_{-1/2}du \frac{dq}{|{\cal M}|} \sum_{ x_n} { A}_j(q) e^{-i p x_n + i q  (x_n + e^{(j)}u)} \nonumber\\
&=&  \int^{1/2}_{-1/2}du  {dq} \delta(p - q ) \tilde{ A}_j(q) e^{ i q   e^{(j)}u}\nonumber\\
&=&  \int^{1/2}_{-1/2}du   \tilde{ A}_{j}(p) e^{ i p   e^{(j)}u}
\nonumber\\
&=&     \tilde{ A}^{(j)}(p) \frac{{\rm sin} (p   e^{(j)}/2)}{(p   e^{(j)}/2)}\label{Str}
\end{eqnarray}
Therefore, in coordinate space we have
\begin{equation}
{\cal A}_\mu(x_1,...,x_\mu,...,x_D) =  \int^{x_\mu+1/2}_{x_\mu-1/2} A_\mu(x_1,...,y_\mu,...,x_D) dy_\mu \,.\label{calA2}
\end{equation}
Here coordinates $x_n$ are taken in lattice units, in which the lattice spacing is equal to unity.
Since the field $A$ is slowly varying, i.e. it almost does not vary on the distance of the order of the lattice spacing, in practical calculations we may substitute to our expressions $A$ instead of $\cal A$. For the real solid state systems the meaning of the slow variation of $A$ is that the magnitude of the external magnetic field is much smaller than thousands of Tesla (the condition that is always fulfilled in practise), and that the wavelength of the external field is much smaller than the lattice spacing. The latter condition corresponds to the wavelengths much larger than several Angstroms, i.e. the above expression for the Weyl symbol (and its extensions to more realistic lattices) cannot not be valid for matter interacting with the X rays. }

\section*{Appendix B. Electric current expressed through the Wigner transformed Green function}

Here we present the proof of the expression for the electric current in the considered lattice model. A slightly different derivation of this expression may be found in \cite{Z2016_1}.
Let us consider the variation of the partition function
\rev{\begin{eqnarray}
\delta {\rm log} \,Z &=&- \frac{1}{Z}
\int D\bar{\psi}D\psi \, {\rm exp}\Big(  {\bf Tr} \hat{W}[\psi,\bar{\psi}] \hat{Q}(p - { A}(i\frac{\partial}{\partial p}))) \Big) \nonumber\\ && \int dx \int \frac{dp}{(2\pi)^D} {\rm Tr} {W}_W[\psi,\bar{\psi}](p,x)\nonumber\\ && \star \partial_{p_k} Q_W(p - { A}(x))\delta{ A}_k(x)\,. \label{dZ2}
\end{eqnarray}}
The total average current \rev{(i.e. the current density integrated over the whole volume of the system)} appears as the response to the variation of $A$:
\begin{eqnarray}
\langle J^k \rangle &=&  -\frac{T}{Z}
\int D\bar{\psi}D\psi \, {\rm exp}\Big(  {\bf Tr} \hat{W}[\psi,\bar{\psi}] \hat{Q}(p - { A}(i\frac{\partial}{\partial p})) \Big) \nonumber\\ && \int d^D x \, \int d^D p\, {\rm Tr} {W}_W[\psi,\bar{\psi}](p,x)  \partial_{p_k} Q_W(p - { A}(x)) \nonumber\\ &=& -T\, \int d^D x \, \int \frac{d^D p}{(2\pi)^D}\, {\rm Tr} \, G_W(p,x)  \partial_{p_k} Q_W(p - { A}(x))\,.\label{J}
\end{eqnarray}
In the presence of periodical boundary conditions the properties of the star product allow to rewrite the last equation in the following way:
\rev{\begin{eqnarray}
\langle J^k \rangle &=& - T\, \int d^D x \, \int \frac{d^D p}{(2\pi)^D}\, {\rm Tr} \, G_W(p,x)\star \partial_{p_k} Q_W(p - { A}(x)) \nonumber\\ &=&   - T\, {\bf Tr} \, \hat{G} \partial_{p_k} {\cal Q}(p - {A}(i\partial_{p})) \,. \label{J2}
\end{eqnarray}}
The last two expressions for the total current are the topological invariants, i.e. the total current is not changed when the system is modified continuously. This may be checked via the consideration of the variation of
Eq. (\ref{J}) corresponding to the variation of $\cal Q$.

\section*{Appendix C. Anomalous QHE. Factorization of topological invariants}

\rv{Let us consider the case, when the fermions are in the presence of Abelian gauge field $A_{i}$. Let us then expand the expression for the electric current density $\langle j^k \rangle$ (the average is understood as an integral over the fermion field) in powers of the external  field strength $A_{ij}$ of $A_i$ and its derivatives. Next, if the coefficient in front of the field strength $A_{ij}$ does not depend on coordinates, then this coefficient is in itself the topological invariant because the integral $\int d^D x A_{ij}$ is not changed, when the gauge potential $A$ is modified smoothly in a finite region of space:}
$$
\delta \int d^D x A_{ij} = \int d^D x \delta A_{ij}  = 0\,.
$$
Here $\delta A \to 0$ at infinity.

This way for $D=3$ we come to the following result for the part of the current proportional to $A_{ij}$:
\begin{equation}
\langle j^{k} \rangle \approx \cor{-} \frac{1}{4\pi}{\cal M}^{ijk} A_{ij}\,.
\end{equation}
Tensor ${\cal M}^{ijk}$ is the topological invariant in momentum space, which was calculated  in \cite{Z2016_1}. The obtained result is
\rev{\begin{eqnarray}
{\cal M}^{ijk} &= & \epsilon^{ijk} {\cal M},\quad  {\cal M} = \int \,{\rm Tr}\, \nu_{} \,d^3p \label{j2d00}\\ \nu_{} & = &  - \frac{i}{3!\,4\pi^2}\,\epsilon_{ijk}\, \Big[  {G}(p) \frac{\partial {G}^{-1}(p)}{\partial p_i} \frac{\partial  {G}( p)}{\partial p_j} \frac{\partial  {G}^{-1}(p)}{\partial p_k} \Big]\,. \nonumber
\end{eqnarray}}
For $D=4$  we obtain
\rev{\begin{eqnarray}
\langle j^{k}\rangle  & \approx & \cor{-} \frac{1}{4\pi^2}\epsilon^{ijkl} {\cal M}_{l} A_{ij}, \label{calM}\\
{\cal M}_l &=& \int_{} \,{\rm Tr}\, \nu_{l} \,d^4p \,.\nonumber \\ \nu_{l} & = & - \frac{i}{3!\,8\pi^2}\,\epsilon_{ijkl}\, \Big[  {G} \frac{\partial {G}^{-1}}{\partial p_i} \frac{\partial  {G}}{\partial p_j} \frac{\partial  {G}^{-1}}{\partial p_k} \Big]  \nonumber\,.
\end{eqnarray}}
Here $G(p)$ is the one - particle Green function of the system without external electric field. It depends on the conserved momentum $p$. \rv{This expression has been used in \cite{Z2016_1} in order to derive the expression for the AQHE in certain topological insulators. The result is expressed through the vectors of the reciprocal lattice. (It also has been obtained earlier in \cite{VolovikTI3D} using another method.) }

\section*{Appendix D. Topological invariant in phase space}

In \cite{Z2016_1} the proof that Eq.(\ref{j2d00}) is the topological invariant was presented. Here we extend this proof to the case of the topological invariant in phase space discussed in the main text of the present paper. \rv{Let us first give the proof for the simplest case of the first term in Eq. (\ref{J32}). We denote it  $J^{(0)}$.  Let us consider an arbitrary variation of the Green function:
${\cal G} \rightarrow {\cal G} + \delta {\cal G}$.
Then using Eq. (\ref{Lbr}), Eq. (\ref{QdG}), and Eq. (\ref{cytr}) we obtain
\begin{eqnarray}
&&\delta J^{(0)k}=
- T\, \delta\int d^D x \, \int \frac{d^D p}{(2\pi)^D}\, {\rm Tr} \, G^{(0)}_W(p,x) \ast \partial_{p_k}  Q^{(0)}_W(p,x)\nonumber\\
&&=- T\,\delta \int_{} \, d^D x \, \int \frac{d^D p}{(2\pi)^D}\, {\rm Tr} G_W* \partial_{p_k}  Q_W \nonumber\\
&&= - T\,\int_{} \,d^D x \, \int \frac{d^D p}{(2\pi)^D}\, {\rm Tr} (\delta G_W * \partial_{p_k}  Q_W+G_W* \partial_{p_k}  \delta Q_W)
\nonumber\\
&&=- T\, \int_{} \, d^D x \, \int \frac{d^D p}{(2\pi)^D}\,{\rm Tr} (-G_W* \delta Q_W * G_W  * \partial_{p_k}  Q_W+G_W* \partial_{p_k}  Q_W)\nonumber\\
&&=- T\, \int_{} \, d^D x \, \int \frac{d^D p}{(2\pi)^D}\,{\rm Tr} (\delta Q_W * \partial_{p_k}  G_W +G_W * \partial_{p_k}  \delta Q_W)\nonumber\\
&&= - T\,\int_{}d^D x \, \int \frac{d^D p}{(2\pi)^D} \, \partial_{p_k} \, {\rm Tr} (\delta Q_W *  G_W)=0\,,
\end{eqnarray}
The last step follows because of the compactness of $p$ space.}

Similarly the expression
for ${\cal M}$ of the main text is changed as follows:
\begin{eqnarray}
&&  \delta {\cal M}   = \rev{-}
  \frac{3\,i\,T}{24  \pi^2\, {\cal V}} \, \int_{} \, {\rm Tr} \Bigl( \Bigl([\delta G_W]\star
  d  Q_W+G_W \star
  d  [\delta Q_W]\Bigr)\nonumber\\&& \star \wedge G_W \star
  d  Q_W \star \wedge G_W \star
  d  Q_W\Bigr)\,d^3 x\nonumber\\&&=\rev{-}\frac{3\,i\,T}{24  \pi^2\, {\cal V}} \int_{} {\rm Tr} \Bigl( \Bigl(-
G_W\star [\delta Q_W]\star G_W \star
  d  Q_W+G_W \star
  d  [\delta Q_W]\Bigr)\nonumber\\ && \star \wedge G_W \star
  d  Q_W \star \wedge G_W \star
  d  Q_W\Bigr)\, d^3 x \nonumber\\&&=
\frac{3\,i\,T}{24  \pi^2\, {\cal V}} \int_{} {\rm Tr} \Bigl(
\Bigl([\delta Q_W] \star [dG_W] +
  d  [\delta Q_W]\star G_W\Bigr)\nonumber\\ && \star \wedge
  d  Q_W \star \wedge
  d  G_W\Bigr)\, d^3 x\nonumber\\&&=
\frac{3\,i\,T}{24  \pi^2\, {\cal V}} \int_{} d \, {\rm Tr} \left(
\Bigl([\delta Q_W] \star G_W] \Bigr)
 \star d  Q_W\star \wedge
  d  G_W\right) \,d^3 x = 0 \nonumber\,.
\end{eqnarray}
\rv{The exterior differentiation is defined here as usual antisymmetric differentiation, which is not affected by the $\star$ product:
\begin{eqnarray}
	dA_W \star\wedge dB_W \star\wedge ... \wedge dC_W = \frac{\partial A_W}{\partial p_{\mu}} \star \frac{\partial B_W}{\partial p_{\nu}} ... \star\frac{\partial C_W}{\partial p_{\rho}} dp_\mu \wedge dp_\nu \wedge ... \wedge dp_\rho
\end{eqnarray}}
%We used here the cyclic property of traces, i.e. that $\int d^3 x d^3 p {\rm Tr} A_W\star B_W = \int d^3 x d^3 p {\rm Tr} B_W\star A_W$. It follows from the corresponding cyclic property of the trace of the operator product:  $\int d^3 x d^3 p {\rm Tr} A_W\star B_W \sim {\bf Tr} \hat{A} \hat{B} = {\bf Tr} \hat{B} \hat{A}$.
%That's {how} we proved that ${\cal M}$ is the topological invariant.

\section*{Appendix E. Ordinary Quantum Hall effect in the $2D$ noninteracting system with a one - particle Hamiltonian}

Here we discuss the Hall conductance that according to the main text is given by $\sigma_{xy} = {\cal N}/2 \pi$ where ${\cal N}$ is the topological invariant in phase space
\rev{\begin{eqnarray}
{\cal N} &= & \frac{T}{3!\,4\pi^2{\cal S}}\, \int \,{\rm Tr}\, d^3p \, d^3x\,   \epsilon_{ijk}\, \Big[  {G}_W(p,x)\ast \frac{\partial {Q}_W(p,x)}{\partial p_i} \nonumber\\ && \ast \frac{\partial  {G}_W( p,x)}{\partial p_j} \ast \frac{\partial  {Q}_W(p,x)}{\partial p_k} \Big]_{A=0}\,.
  \label{calM2d23}
\end{eqnarray}}
Our aim is to derive starting from Eq. (\ref{calM2d23}) the following representation for $\cal N$ in terms of the Green function written in momentum representation:
\rev{\begin{eqnarray}
{\cal N} &=&  \frac{T \,(2\pi)^3}{3!\,4\pi^2{\cal S}}\, \int \,{\rm Tr}\, d^3p^{(1)} \, d^3p^{(2)}\, d^3 p^{(3)} \, d^3 p^{(4)}\,       \epsilon_{ijk}\,\nonumber\\&& \Big[  {G}(p^{(1)},p^{(2)})\Big( [\partial_{p^{(2)}_i} + \partial_{p^{(3)}_i}] Q(p^{(2)},p^{(3)})\Big)\nonumber\\ &&  \Big( [\partial_{p^{(3)}_j} + \partial_{p^{(4)}_j}]  G(p^{(3)},p^{(4)}) \Big)\nonumber\\ && \Big( [\partial_{p^{(4)}_k} + \partial_{p^{(1)}_k}] Q(p^{(4)},p^{(1)})\Big) \Big]_{A=0}\,.
  \label{calM2d23P}
\end{eqnarray}}
In order to derive this representation we first represent   $\partial_{p^i}G_W$ as follows:
{\begin{eqnarray}
	&&\frac{\partial}{\partial p^i}G_W(p,x)=\frac{\partial}{\partial p^i}\int d^3P~e^{iPx}G\Big(p+\frac{P}{2},p-\frac{P}{2}\Big)
	\\
	&&=\int d^3P~e^{iPx}\Big(\frac{\partial}{\partial K_1^i }+\frac{\partial}{\partial {K_2^i} })G(K_1,K_2)\Big|_{K_1=p+\frac{P}{2}}^{K_2=p-\frac{P}{2}}\,,\nonumber
\end{eqnarray}}
where we used
$$
	\frac{\partial}{\partial {p^i} }
	=\frac{\partial K_1^j}{\partial {p^i} }\frac{\partial}{\partial {K_1^j} }+\frac{\partial K_2^j}{\partial {p^i} }\frac{\partial}{\partial {K_2^j} }	=\frac{\partial}{\partial {K_1^i} }+\frac{\partial}{\partial {K_2^i} }\,.
$$
We denote
{
$$
	G'_i\Big(p+\frac{P}{2},p-\frac{P}{2}\Big)=\Big(\frac{\partial}{\partial {K_1^i} }+\frac{\partial}{\partial {K_2^i} }\Big)G(K_1,K_2)\Big|_{K_1=p+\frac{P}{2}}^{K_2=p-\frac{P}{2}}\,,
$$
for convenience. So $\partial_{p^i}G_W$ becomes
\begin{eqnarray}
	\partial_{p^i}G_W=(G'_i)_W\,.
\end{eqnarray}}
Next we can compute $G_W*\partial_{p^i}Q_W$.
$$
	Q_W*\partial_{p^i}G_W=(QG'_i)_W=\int d^3P~e^{iPx}(QG_i')\Big(p+\frac{P}{2},p-\frac{P}{2}\Big)\,,
$$
in which
$$
	(QG_i')\Big(p+\frac{P}{2},p-\frac{P}{2}\Big) =\int\frac{d^3p'}{(2\pi)^3}~G\Big(p+\frac{P}{2},p'\Big)$$$$\Big(\frac{\partial}{\partial {p'^i} }+\frac{\partial}{\partial {K_2^i} }\Big)Q(p',K_2)\Big|_{K_2=p-\frac{P}{2}}\,.
$$

Using the associativity of $*$-product, similarly we have
{
\begin{eqnarray}
	&&{Q}_W(p,x)\ast \frac{\partial {G}_W(p,x)}{\partial p_i} \ast \frac{\partial  {Q}_W( p,x)}{\partial p_j} \ast \frac{\partial  {G}_W(p,x)}{\partial p_k} \nonumber
	\\
	&&=(QG'_iQ'_jG'_k)_W\nonumber
	\\
	&&=\int d^3P~e^{iPx}(QG'_iQ'_jG'_k)\Big(p+\frac{P}{2},p-\frac{P}{2}\Big)\,.
	\nonumber
\end{eqnarray}}
Integrating over $x$ and $P$, inserting the completeness relation $1=\int\frac{d^3p_i}{(2\pi)^3}| p_i \rangle \langle p_i |$, we transform the expression for  ${\cal N}$ into the form of
Eq.(\ref{calM2d23P}).

Our next purpose is to bring Eq. (\ref{calM2d23P}) to the conventional expression for the Hall conductance in the case, when
the non - interacting charged fermions with Hamiltonian ${\cal H}(p_x,p_y)$ are in the presence of constant external magnetic field ${\cal B}$ orthogonal to the plane of the given $2D$ system.
We use the gauge, in which the gauge potential is
$$
B_x = 0, \quad B_y = {\cal B} x\,.
$$
External electric field ${\it E}_y$ corresponding to the gauge potential $A$ is directed along the axis $y$.

Function $Q(p^{(1)},p^{(2)})$ in momentum space has the form:
$$
Q(p^{(1)},p^{(2)}) = \langle p^{(1)}| \hat{Q} | p^{(2)}\rangle = \zz{\Big(}\delta^{(3)} (p^{(1)}-p^{(2)}) i \omega^{(1)} $$$$ - \langle {\bf p}^{(1)}| {\cal H} | {\bf p}^{(2)}\rangle \zz{\Big)} \delta(\omega^{(1)}-\omega^{(2)})
$$
where $p = (p_1,p_2,p_3) = ({\bf p},\omega)$. At the same time
$$
G(p^{(1)},p^{(2)}) = \sum_{n} \frac{1}{i\omega^{(1)} - {\cal E}_n} \langle {\bf p}^{(1)}| n \rangle \langle n | {\bf p}^{(2)}\rangle \delta(\omega^{(1)}-\omega^{(2)})\,.
$$
This way we obtain:
\rev{\begin{eqnarray}
&&{\cal N} =  - \frac{i\,(2\pi)^2}{8\pi^2{\cal S}}\,\sum_{n,k} \int \,d \omega  d^2{\bf p}^{(1)} \, d^2{\bf p}^{(2)}\, d^2 {\bf p}^{(3)} \, d^2 {\bf p}^{(4)}\,       \nonumber\\ && \epsilon_{ij}\,{\rm Tr}\,\Big[  \frac{1}{(i\omega^{} - {\cal E}_n)^2} \langle {\bf p}^{(1)}| n \rangle \langle n | {\bf p}^{(2)}\rangle \nonumber\\ && \Big( [\partial_{p^{(2)}_i} + \partial_{p^{(3)}_i}] \langle {\bf p}^{(2)}| {\cal H} | {\bf p}^{(3)}\rangle\Big)  \nonumber\\&& \frac{1}{(i\omega^{} - {\cal E}_k)} \langle {\bf p}^{(3)}| k \rangle \langle k | {\bf p}^{(4)}\rangle  \Big( [\partial_{p^{(4)}_j} + \partial_{p^{(1)}_j}]  \langle {\bf p}^{(4)}| {\cal H} | {\bf p}^{(1)}\rangle \Big) \Big]_{A=0}\,.\nonumber
\end{eqnarray}}
One may represent
$$
[\partial_{p^{(4)}_j} + \partial_{p^{(1)}_j}]  \langle {\bf p}^{(4)}| {\cal H} | {\bf p}^{(1)}\rangle% =\langle  {\bf p}^{(4)} | \frac{\partial }{\partial {\hat{p}^{i}}}{\cal H} | {\bf p}^{(1)} \rangle
= i   \langle {\bf p}^{(4)}| {\cal H} {\hat x}_j -{\hat x}_j{\cal H}   | {\bf p}^{(1)}\rangle$$$$
 = i \langle {\bf p}^{(4)}| [{\cal H}, {\hat x}_j]| {\bf p}^{(1)}\rangle\,.
$$
By operator $\hat x$ we understand $i\partial_{p}$ acting on the the wavefunction written in momentum representation:
$$
\hat{x}_i \Psi(p) = \langle p|\hat{x}_i |\Psi\rangle = i\partial_p \langle p|\Psi\rangle = i \partial_p \Psi(p)\,.
$$
Then, for example, $$\hat{x}_i \delta(q-p) = \langle p|\hat{x}_i |q\rangle = i\partial_p \langle p|q \rangle = i \partial_p \delta(p-q) = -i \partial_q \langle p|q \rangle \,.$$
Therefore, we can write
$$
\hat{x}_i |p\rangle = -i\partial_p |p \rangle\,.
$$
Notice, that the sign minus here is counter - intuitive because the operator $\hat x$ is typically associated with $+i\partial_p$. We should remember, however, that with this latter representation the derivative acts on $p$ in the bra - vector $\langle p| $ rather than on $p$ in $| p \rangle$. Above we have shown that the sign is changed when the derivative is transmitted to $p$ of $| p \rangle$.

Thus we have
\rev{\begin{eqnarray}
{\cal N} &=&  \frac{i\,(2\pi)^2}{8\pi^2{\cal S}}\,\sum_{n,k} \int \,d \omega  \, \epsilon_{ij}\,\Big[  \frac{1}{(i\omega^{} - {\cal E}_n)^2}  \langle n| [{\cal H}, {\hat x}_i] | k \rangle \nonumber\\ && \frac{1}{(i\omega^{} - {\cal E}_k)}  \langle k | [{\cal H}, {\hat x}_j] | n \rangle  \Big]_{A=0}\nonumber\\
&=& - \frac{2i\,(2\pi)^3}{8\pi^2{\cal S}}\,\sum_{n,k}   \, \epsilon_{ij}\,\Big[  \frac{1}{({\cal E}_k - {\cal E}_n)^2} \nonumber\\ && \langle n| [{\cal H}, {\hat x}_i] | k \rangle    \langle k | [{\cal H}, {\hat x}_j] | n \rangle  \Big]_{A=0}\theta(-{\cal E}_n)\theta({\cal E}_k)\,.\label{sigmaHH}
\end{eqnarray}}
The last expression is just the conventional expression for the Hall conductance (\rev{multiplied by $2\pi$}) for the given system \cite{Tong:2016kpv}.

\section*{Appendix F. Hall conductance in the $2+1$ D systems from the Kubo formula}

According to our previous considerations, in the particular case, when the external gauge field $A$ represents constant electric field $E_k$, we get $A_{3k} =  -i E_k$. Therefore, Eq. (\ref{JA2D}) reads
\begin{equation}
\langle j^{k} \rangle \approx \cor{-}\frac{{\cal N}}{2\pi} \epsilon^{3kj} E_{j}\,.\label{JA2D}
\end{equation}
Here ${\cal N}$ is the topological invariant in phase space
\begin{eqnarray}
{\cal N} &= & \rev{ \frac{T}{\cal S} \int \,{\rm Tr}\, \nu_{} \,d^3p \, d^3x}\label{j2d}\,,\nonumber \\ \nu_{} & = &   \frac{1}{3!\,4\pi^2}\,\epsilon_{ijk}\, \Big[  {G}_W(p,x)\ast \frac{\partial {Q}_W(p,x)}{\partial p_i} \nonumber\\ && \ast \frac{\partial  {G}_W( p,x)}{\partial p_j} \ast \frac{\partial  {Q}_W(p,x)}{\partial p_k} \Big]_{A=0} \,. \label{calM2d2H}
\end{eqnarray}

Above we demonstrated that this expression is reduced to Eq. (\ref{sigmaHH}). Here for completeness we present the standard derivation of Eq. (\ref{sigmaHH}).
We will use the Kubo formula for Hall conductance (see, for example, \cite{Tong:2016kpv}). Let us start from the consideration of one fermionic particle that may occupy the states enumerated by numbers $n = 0, 1, ... $.  We have
\be \label{KB}
	\sigma_{xy}&=& \frac{1}{\omega {\cal S}}\int_{0}^{\infty} dt ~e^{i\omega t}\langle 0 | [\hat{J}_y(0), \hat{J}_x(t)]|0 \rangle\,.
\ee
Here  $$ \hat{J}_x(t)=\hat{V}^{-1}(t)\hat{J}_x(0)\hat{V}(t)=e^{i{\cal H}_0t}\hat{J}_x(0)e^{-i{\cal H}_0t}\,$$
is the total electric current in the Heisenberg picture, while $V(t)$ is the time evolution operator. Averaging is over the the occupied state $|0 \rangle$.
Substituting it into the Kubo formula of Eq. (\ref{KB}), inserting a complete set of energy eigenstates, we get
\be
	\sigma_{xy}&=& \frac{1}{\omega  {\cal S}} \int_{0}^{\infty} dt ~e^{i\omega t}\sum_n\Big(\langle 0 | \hat{J}_y |n\rangle \langle n|\hat{J}_x| 0 \rangle e^{i({\cal E}_n-{\cal E}_{0})t}\nonumber\\ &&-\langle 0 | \hat{J}_x |n\rangle \langle n|\hat{J}_y| 0 \rangle e^{-i({\cal E}_n-{\cal E}_{0})t}\Big),
\ee
where we write $J_i=J_i(0)$ for simplicity, and the $n=0$ term vanishes.  ${\cal E}_n$ is the energy of state $|n \rangle$.
Next we integrate over time,
and get
\be \nonumber
	\sigma_{xy}&=&\frac{1}{i\omega  {\cal S}}  \sum_{n\ne 0} \Big(\frac{\langle 0 | \hat{J}_y |n\rangle \langle n|\hat{J}_x| 0 \rangle}{\omega+i\epsilon+{\cal E}_n-{\cal E}_{0}}-\frac{\langle 0 | \hat{J}_x |n\rangle \langle n|\hat{J}_y| 0 \rangle}{\omega+i\epsilon-({\cal E}_n-{\cal E}_{0})}\Big).
\ee
To get the limit $\omega \to 0$ corresponding to constant background field, we expand the above expression in powers of $\omega$
\be
	\sigma_{xy}&=&\frac{1}{i\omega  {\cal S}} \sum_{n\ne 0} \Big( \langle 0 | \hat{J}_y |n\rangle \langle n|\hat{J}_x| 0 \rangle \Big(\frac{1}{{\cal E}_n-{\cal E}_{0}}-\frac{\omega}{({\cal E}_n-{\cal E}_{0})^2}+...\Big)\nonumber
	\\
	&&+\langle 0 | \hat{J}_x |n\rangle \langle n|\hat{J}_y| 0 \rangle \Big(\frac{1}{{\cal E}_n-{\cal E}_{0}}+\frac{\omega}{({\cal E}_n-{\cal E}_{0})^2}+...\Big)\Big)  \,.
\ee
For the Hall effect
	$$\sigma_{xy}=-\sigma_{yx}\,,$$
which leads to vanishing of the first order term.
In the second order, the Hall conductance becomes
\be
	\sigma_{xy}&=&\frac{i} {\cal S}\sum_{n\ne 0}\frac{\langle 0 | \hat{J}_y |n\rangle \langle n|\hat{J}_x| 0 \rangle-\langle 0 | \hat{J}_x |n\rangle \langle n| \hat{J}_y | 0 \rangle}{({\cal E}_n-{\cal E}_{0})^2}\,.
\ee
Electric current may be written as
\be
\hat{J}_i=\frac{1}{i}[\hat{x}_i,{\cal H}]\,,
\ee
Thus the Hall conductance of this quantum - mechanical system may be  written as
\be
	\sigma_{xy}&=&\frac{1}{i {\cal S}}\sum_{n\ne0}\frac{\langle 0 | [\hat{y},{\cal H}] |n\rangle\langle n|  [\hat{x},{\cal H}] | 0\rangle-\langle 0 |  [\hat{x},{\cal H}] |n\rangle\langle n| [\hat{y},{\cal H}] | 0\rangle}{({\cal E}_n-{\cal E}_0)^2}\,.\nonumber
\ee
This equation represents the result of the second order in the perturbation expansion that corresponds to the transition between the states: from $|0\rangle $ to $|n \rangle$, and back from $|n \rangle $ to $|0 \rangle$.

The next step is the consideration of the second quantized system. Fermi statistics provides, that in vacuum all one particle states with energy (counted from the Fermi level) ${\cal E}_k < 0$ are occupied. The Hall conductance in the second order of perturbation theory is given by the contributions of the transitions from the occupied one - particle states to the vacant ones and the corresponding returns. The final answer reads:
\be
&&	\sigma_{xy}=\frac{i} {\cal S} \sum_k \sum_{n\ne k}\theta(-{\cal E}_k)\theta({\cal E}_n)\nonumber\\ &&\frac{\langle k | [\hat{x},{\cal H}] |n\rangle\langle n|  [\hat{y},{\cal H}] | k \rangle-\langle k |  [\hat{y},{\cal H}] |n\rangle\langle n| [\hat{x},{\cal H}] | k\rangle}{({\cal E}_n-{\cal E}_k)^2}\,.\label{sigmaH}
\ee
This expression gives rise to the Hall conductance of the form \rev{$\sigma_{yx} = {\cal N}/2 \pi$} where ${\cal N}$ is the topological invariant given by Eq. (\ref{sigmaHH}). The latter expression, in turn, was derived from the phase space topological invariant of Eq. (\ref{calM2d2H}). Thus the consideration of the present section gives an alternative proof of Eq. (\ref{calM2d2H}) in the given particular case.

\section*{Appendix G. Calculation of the Hall conductance for the noninteracting 2D system in the presence of constant magnetic field}

The average Hall conductivity may be represented as ${\cal N}/(2\pi)$, where
\rev{\begin{eqnarray}
{\cal N} &=& - \frac{2i\,(2\pi)^3}{8\pi^2  {\cal S}}\,\sum_{n,k} \theta({\cal E}_k)\theta(-{\cal E}_n)  \, \epsilon_{ij}\,\nonumber\\&&\Big[  \frac{1}{({\cal E}_k - {\cal E}_n)^2}  \langle n| [{\cal H}, {\hat x}_i] | k \rangle    \langle k | [{\cal H}, {\hat x}_j] | n \rangle  \Big]_{A=0}\,.\label{sigmaHH__}
\end{eqnarray}}
Following \cite{QHEB} (see also \cite{Maraner}) in order to calculate the value of $\cal N$ we decompose the coordinates $x_1=x, x_2=y $ as follows:
$$
\hat{x}_1 = -\frac{\hat{p}_y - {\cal B}x}{\cal B} + \hat{X}_1 = \hat{\xi}_1 + \hat{X}_1,$$ $$ \hat{x}_2 = \frac{\hat{p}_x}{\cal B} + \hat{X}_2= \hat{\xi}_2 + \hat{X}_2\,.
$$
The commutation relations follow:
$$
[\hat{\xi}_1,\hat{\xi}_2] = \frac{i}{\cal B}\,, \quad [\hat{X}_1,\hat{X}_2] = - \frac{i}{\cal B}\,,
$$
$$
[{\cal H}, \xi_1] = -i  \frac{\partial}{\partial p_x} {\cal H}\,, \quad [{\cal H}, \xi_2] =  -i \frac{\partial}{\partial p_y} {\cal H}\,,
$$
$$
[{\cal H}, \hat{X}_1] =  [{\cal H}, \hat{X}_2] =  0\,.
$$
Here we use that the Hamiltonian contains the following dependence on $x$:
$$
{\cal H}(\hat{p}_x, \hat{p}_y - {\cal B} x)\,
$$
and assume $\frac{\partial^2}{\partial p_x \partial p_y}{\cal H}=0$.
We use those relations to obtain:
\rev{\begin{eqnarray}
{\cal S N} &=&  - \frac{2i\,(2\pi)^3}{8\pi^2}\,\sum_{n,k}   \,\Big[  \frac{1}{({\cal E}_k - {\cal E}_n)^2}  \langle n| [{\cal H}, {\hat \xi}_i] | k \rangle    \langle k | [{\cal H}, {\hat \xi}_j] | n \rangle  \Big]_{A=0}\nonumber\\&&\, \epsilon_{ij} \, \theta(-{\cal E}_n)\theta({\cal E}_k)
\nonumber\\
&=&  \frac{2i\,(2\pi)^3}{8\pi^2}\,\sum_{n,k}   \, \epsilon_{ij}\,\Big[  \langle n|  {\hat \xi}_i | k \rangle    \langle k |  {\hat \xi}_j | n \rangle  \Big]_{A=0}\theta(-{\cal E}_n)\theta({\cal E}_k)
\nonumber\\
&=&  \frac{2i\,(2\pi)^3}{8\pi^2}\,\sum_{n}   \,\Big[  \langle n|  [{\hat \xi}_1,  {\hat \xi}_2 ]| n \rangle  \Big]_{A=0}\theta(-{\cal E}_n)
\nonumber\\
&=&  -\frac{2\,(2\pi)^3}{8\pi^2 {\cal B}}\,\sum_{n}   \,   \langle n|   n \rangle  \theta(-{\cal E}_n)\,.\label{QHEB}
\end{eqnarray}}

Momentum $p_y$ is a good quantum number, and it enumerates the eigenstates of the Hamiltonian:
$$
{\cal H} |n\rangle = {\cal H}(\hat{p}_x, p_y - {\cal B} x)|p_y, m\rangle = {\cal E}_{m}(p_y)|p_y, m\rangle , \, m\in Z\,.
$$
We assume that the size of the system is $L\times L$.
This gives
\rev{\begin{eqnarray}
{\cal S N} &=& - {(2\pi)}\,\sum_{m}\int \frac{dp_y L}{2\pi}  \, \frac{1}{\cal B} \theta(-{\cal E}_m(p_y))\,.
  \label{calM2d232}
\end{eqnarray}}
$\langle x \rangle = p_y/{\cal B}$ plays the role of the center of orbit, and this center should belong to the interval $(-L/2, L/2)$. This gives
\rev{\begin{eqnarray}
{\cal N}
&=&  N  \, {\rm sign}(-{\cal B})\,,
  \label{calM2d233}
\end{eqnarray}}
where $N$ is the number of the occupied branches of spectrum. This way we came to the conventional expression for the Hall conductance of the fermionic system in the presence of constant magnetic field and constant electric field.

\section*{Appendix H. Properties of Moyal (star) product}
\rv{In this Appendix, we list and prove the identities of Wigner - Weyl formalism used throughout the text of the paper. Wigner transformation of a function $B(p_1,p_2)$ (where  $p_1, p_2 \in {\cal M}$) is defined here as
\begin{equation} \begin{aligned}
{B}_W(x,p) \equiv \int_{q \in {\cal M}}  dq e^{ix q} B({p+q/2},{p-q/2}) \label{GWxAH0}
\end{aligned}\,.
\end{equation}
Identifying $B(p_1,p_2)$ with the matrix elements of an operator $\hat B$, we come to the definition of the Weyl symbol of operator
(we denote it by $B_W$):
 \begin{equation} \begin{aligned}
{B}_W(x,p) \equiv \int_{q \in {\cal M}} dq e^{ix q} \langle {p+q/2}| \hat{B} |{p-q/2}\rangle \label{GWxAH}
\end{aligned}\,.
\end{equation}
Here integral is over the Brillouin zone. Moyal product of the two functions in phase space $f(x,p)$ and $g(x,p)$ is defined as 	$$f(x,p)\star g(x,p) \equiv f(x,p) e^{\frac{i}{2} \left( \overleftarrow{\partial}_{x}\overrightarrow{\partial_p}-\overleftarrow{\partial_p}\overrightarrow{\partial}_{x}\right )}g(x,p)$$
Let us consider the case, when operators $\hat A$ and $\hat B$ are almost diagonal, i.e. $\langle {p+q/2}| \hat{A} |{p-q/2}\rangle$ and $\langle {p+q/2}| \hat{B} |{p-q/2}\rangle$ may be nonzero for arbitrary $p$ and small $q$ (compared to the size of momentum space). This occurs when the variation of $A_W(x,p)$ (and $B_W(x,p)$) as a function on $x$ may be neglected on the distances of the order of the lattice spacing. Below we assume that the considered operators satisfy this requirement. Then the following  expression follows
\begin{eqnarray}\label{Moyal2}
(\hat{A}\hat{B})_W(x,p) = 	A_W(x,p)\star B_W(x,p) = A_W(x,p) e^{\frac{i}{2} \left( \overleftarrow{\partial}_{x}\overrightarrow{\partial_p}-\overleftarrow{\partial_p}\overrightarrow{\partial}_{x}\right )}B_W(x,p)\,.
\end{eqnarray}
The proof is given in \cite{SZ2018}. We repeat it here briefly:
\begin{equation}\begin{aligned}
&(AB)_W(x,p )=
 \int_{{\cal M}} d{P} \int_{\cal M} d{R}
 	e^{i x P}
 		\Bra{p+\frac{P}{2}} \hat{A} \Ket{R}
		\Bra{R} \hat{B} \Ket{p-\frac{P}{2}}\\
&=\frac{1}{2^D}\int_{{\cal M}} d P \int_{K/2 \in {\cal M}}dK
	e^{i x P}
		\Bra{p+\frac{P}{2}} \hat{A} \Ket{p-\frac{K}{2}}
		\Bra{p-\frac{K}{2}}\hat{B} \Ket{p-\frac{P}{2}}\\
&= \frac{2^D}{2^D }\int_{{\cal M}} d q d k e^{i x (q+ k)}
	\Bra{p+\frac{q}{2}+\frac{k}{2}} \hat{A} \Ket{p-\frac{q}{2}+\frac{k}{2}}
	\Bra{p-\frac{q}{2}+\frac{k}{2}}\hat{B} \Ket{p-\frac{q}{2}-\frac{k}{2}}\\
&= \int_{{\cal M}} d q d k
	\Big[  e^{i x  q}
		\Bra{p+\frac{q}{2}} \hat{A} \Ket{p-\frac{q}{2}}
	\Big]
	e^{\frac{k}{2}\overleftarrow{\partial}_p-\frac{q}{2}\vec{\partial}_p}
	\Big[  e^{i x   k}
		\Bra{p+\frac{ k}{2}}\hat{B} \Ket{p-\frac{ k}{2}}
	\Big]\\
&= \Big[ \int_{{\cal M}} dq  e^{i x  q}
		\Bra{p+\frac{q}{2}} \hat{A} \Ket{p-\frac{q}{2}}
	\Big]
	e^{\frac{i}{2} (- \overleftarrow{\partial}_p\vec{\partial}_x+\overleftarrow{\partial}_{x}\vec{\partial}_p )}
	\Big[ \int_{{\cal M}} dk e^{i x   k}
		\Bra{p+\frac{k}{2}}\hat{B} \Ket{p-\frac{k}{2}}
	\Big]
\label{Z}
\end{aligned}
\end{equation}
Here the bra- and ket- vectors in momentum space are defined modulo vectors of reciprocal lattice. In the second line we change variables
$$
	P = q+k , \quad K = q- k
$$
$$
	q = \frac{P+K}{2}, \quad  k =\frac{P-K}2
$$
with the Jacobian
$$
J = \left|\begin{array}{cc} 1 & 1 \\
-1 & 1 \end{array} \right| = 2^D
$$
This results in the factor ${2^{D}}$ in the third line. Here $D$ is the dimension of space. The transition between the second and the third lines of Eq. (\ref{Z}) requires that the operators are almost diagonal. This allows to substitute the region of the values of $q$ and $k$ (that corresponds to $P,K/2 \in {\cal M}$) by ${\cal M}\otimes {\cal M}$.}

\rv{In the present paper we apply the Wigner - Weyl technique to the lattice Dirac operator (the inverse Green function) and the Green function $G(p,q)$ itself (to be considered as matrix elements of an operator $\hat G$: $G(p+q/2,p-q/2) =\langle {p+q/2}| \hat{G} |{p-q/2}\rangle$. Both are almost diagonal if the external electromagnetic field varies slowly, i.e. if its variation on the distance of the order of lattice spacing may be neglected. This occurs for the magnitudes of external magnetic field much smaller than thousands Tesla, and for the wavelengths much larger than $1$ Angstrom.
One has
\begin{eqnarray}
	G_W(x,p)\star Q_W(x,p)=(\hat{G}\hat{Q})_W(x,p)&=&\int_{q\in {\cal M}} dq e^{ix q} \langle {p+q/2}| \hat{G} \hat{Q} |{p-q/2}\rangle \nonumber
	\\
	&=&\int_{q \in {\cal M}} dq e^{ix q} \delta(q)=1\,.
\end{eqnarray}
Thus we have
\begin{eqnarray}
	Q_W(x,p)\star G_W(x,p)=1\,.
\end{eqnarray}}

\rv{From the definition the associativity of the Moyal product follows: $[[A_W\star B_W]\star C_W]=[A_W\star [B_W\star C_W]]$.
The proof is as follows:
\begin{eqnarray}
	&&[[A_W(x,p)\star B_W(x,p)]\star C_W(x,p)]\nonumber
	\\
	&=&[(AB)_W(x,p) \star B_W(x,p)]\nonumber
	\\
	&=&(ABC)_W(x,p)\nonumber
	\\
	&=&[A_W(x,p)\star (BC)_W(x,p)]\nonumber
	\\
	&=&[A_W(x,p)\star [B_W(x,p)\star C_W(x,p)]]\,.
\end{eqnarray}}

\rv{The Leibnitz product rule is valid for the Weyl symbols of considered operators.
Let $d$ be a differentiation operator $d=d x^{\mu} \frac{\partial}{\partial x^{\mu}}+d p_{\mu} \frac{\partial}{\partial p_{\mu}}$, then
\begin{eqnarray}\label{Lbr}
d (A_W(x,p)\star B_W(x,p))=d A_W(x,p)\star B_W(x,p)+A_W(x,p)\star d B_W(x,p)\,.
\end{eqnarray}
The proof is straightforward.
First, notice that
\begin{eqnarray}
	d (A_W\frac{i}{2}\left( \overleftarrow{\partial}_{x}\overrightarrow{\partial_p}-\overleftarrow{\partial_p}\overrightarrow{\partial}_{x}\right )B_W)
	&=&d A_W(x,p)\frac{i}{2}\left( \overleftarrow{\partial}_{x}\overrightarrow{\partial_p}-\overleftarrow{\partial_p}\overrightarrow{\partial}_{x}\right ) B_W(x,p)\nonumber
	\\
	&&+A_W(x,p)\frac{i}{2}\left( \overleftarrow{\partial}_{x}\overrightarrow{\partial_p}-\overleftarrow{\partial_p}\overrightarrow{\partial}_{x}\right ) d B_W(x,p)\,.
\end{eqnarray}
Then we expand the star
\begin{eqnarray}
d (A_W(x,p)\star B_W(x,p))	
&=&d  \sum_n(A_W\frac{1}{n!}(\frac{i}{2})^n\left( \overleftarrow{\partial}_{x}\overrightarrow{\partial_p}-\overleftarrow{\partial_p}\overrightarrow{\partial}_{x}\right )^nB_W)\nonumber\\
	&=&d A_W(x,p)\sum_n\frac{1}{n!}(\frac{i}{2})^n\left( \overleftarrow{\partial}_{x}\overrightarrow{\partial_p}-\overleftarrow{\partial_p}\overrightarrow{\partial}_{x}\right )^n B_W(x,p)\nonumber
	\\
	&&+\sum_nA_W(x,p)\frac{1}{n!}(\frac{i}{2})^n\left( \overleftarrow{\partial}_{x}\overrightarrow{\partial_p}-\overleftarrow{\partial_p}\overrightarrow{\partial}_{x}\right )^n d B_W(x,p)\nonumber\\
	&=&
	d A_W(x,p)\star B_W(x,p)+A_W(x,p)\star d B_W(x,p)\,.
\end{eqnarray}
From the Leibnitz product rule and the Groenewold equation we derive
\begin{eqnarray}\label{QdG}
	&&Q_W \star dG_W=-dQ_W \star G_W,\nonumber\\
	&&d G_W= -G_W\star dQ_W \star G_W\,.
\end{eqnarray}}

\rv{Since the Weyl symbols of the considered operators vary slowly as functions of coordinates, we may change everywhere the sum over the lattice points $x$ to the integral over $x$. Under this condition it follows from the definition of the Weyl symbol of an operator that
$$
\int d^D p \Bra{p} \hat{A} \Ket{p} = \int d^Dx \frac{d^Dp}{(2\pi)^D} A_W(x,p)
$$
The integral of a Moyal product over phase space has the following commutativity property
\begin{eqnarray}\label{cytr}
	\int d^Dx\,\frac{d^Dp}{(2\pi)^D}\,{\rm Tr} \, A_W(x,p) \star B_W(x,p)=  \int d^Dx\,\frac{d^Dp}{(2\pi)^D}\,{\rm Tr} \, B_W(x,p) \star A_W(x,p)
	\,.
\end{eqnarray}
We prove it as follows
\begin{eqnarray}
	\int d^Dx\,\frac{d^Dp}{(2\pi)^D}\,{\rm Tr} \, A_W(x,p) \star B_W(x,p)\nonumber&=& \int_{} \, d^Dx\frac{d^Dp}{(2\pi)^D} {\rm Tr} (AB)_W(x,p) \nonumber\\
	&=&  \int_{} \, d^Dp \, {\rm Tr} (AB)(p,p)\nonumber\\
	&=&  \int_{} \, d^Dp\,d^Dp' \,{\rm Tr} A(p,p')B(p',p)\nonumber\\
	&=&  \int_{} \, d^Dp\,d^Dp' \, {\rm Tr} B(p',p)A(p,p')\nonumber\\
	&=&  \int_{} \, d^Dp' \,{\rm Tr} (BA)(p',p')\nonumber\\
	&=&  \int d^Dx\,\frac{d^Dp}{(2\pi)^D}\,{\rm Tr} \, B_W(x,p) \star A_W(x,p)
	\,.
\end{eqnarray}}

\end{document}